\documentclass[preprint,12pt,authoryear]{elsarticle}

\usepackage{amsmath,amssymb}
\usepackage{graphicx}
\usepackage{booktabs}
\usepackage{multirow}
\usepackage[utf8]{inputenc}
\usepackage[T1]{fontenc}
\usepackage{lineno}
\usepackage[colorlinks=true,allcolors=blue]{hyperref}

\newenvironment{tabnote}{\par\vspace{2pt}\footnotesize}{\par}

\journal{Icarus}

\begin{document}

\begin{frontmatter}

\title{Beyond gas dwarfs and water worlds: hybrid envelopes among the
M-dwarf sub-Neptunes characterised with JWST}

\author{J.~F. Pag\'es Navarrete\corref{cor1}}
\ead{jfpages@gmail.com}
\cortext[cor1]{Corresponding author. ORCID: 0009-0006-7910-4276}
\address{Independent Researcher, M\'alaga, Spain}

\begin{abstract}
Sub-Neptunes orbiting M dwarfs are commonly interpreted either as rocky
cores beneath primordial H/He envelopes (gas dwarfs) or as volatile-rich
bodies formed beyond the water-ice line (water worlds). Because the two are
degenerate in mass and radius once envelope evolution is treated
self-consistently, the atmospheric mean molecular weight $\mu$ is the more
direct discriminant. We compile the eight M-dwarf sub-Neptunes with
published JWST transmission spectra and a constraint on $\mu$, grade each
constraint by the quality of the underlying evidence, and combine it with
the restricted Jeans parameter and with energy-limited escape integrated
over the system lifetime, using XUV histories from M-dwarf activity--age
relations. Planets are classified by what their present composition implies
about their origin. At the $\mu = 3.8$\,amu threshold adopted in the
gas-dwarf literature, four planets fit neither end-member: their envelopes
($3.8 < \mu \le 18$\,amu) are too heavy for primordial H/He, yet neither
their composition nor their escape history requires a volatile-rich origin.
Two of these assignments rest on molecular detections (LP~791-18~c and
TOI-270~d) and two on bounded or conditional metallicities. One planet,
LHS~1140~b, requires a volatile-rich origin; one, TOI-776~c, is tentatively
a gas dwarf; two remain ambiguous. We propose the term hybrid worlds for the
intermediate class, whose envelopes can be produced by magma-ocean
equilibration, supercritical miscibility, preferential hydrogen loss or
accretion interior to the ice line. The count depends on the adopted
threshold (three planets at 5\,amu, one at 10\,amu), but no threshold leaves
both end-member classes physically consistent. The sample is set by JWST
target selection, so these fractions are not occurrence rates; a homogeneous
survey testing whether $\mu$ is bimodal would test the proposed class
directly.
\end{abstract}


\end{frontmatter}

\section{Introduction}
\label{sec:1}

Sub-Neptunes have no solar system analogue and no unique interior solution. A planet of
2.5 $R_\oplus$ and 8 $M_\oplus$ is reproduced equally well by a rocky core beneath a few per cent of hydrogen
and helium by mass, or by a volatile-rich interior with no primordial envelope at all \citep{rogersseager10,valencia07}. Mass and radius alone therefore cannot determine what these
planets are made of, and the difficulty is sharpest for the temperate sub-Neptunes of M dwarfs,
which are both the most abundant small planets known and the most accessible to transmission
spectroscopy.

Two models have framed the interpretation of this population. In the gas-dwarf picture,
super-Earths and sub-Neptunes are a single population of rocky cores that accreted H/He from the
protoplanetary disc; XUV photoevaporation and core-powered mass loss subsequently strip the
lower-mass and more irradiated planets, carving the radius valley \citep{owenwu13,owenwu17,ginzburg18,gupta19}. Hierarchical inference on the Kepler sample recovers
initial conditions consistent with this picture: cores of terrestrial rock-iron composition,
formed in a water-poor environment, most of which accreted envelopes of a few per cent by mass
\citep{rogersowen21}. In the water-world picture, \citet{luque22} argued from bulk densities
that M-dwarf planets separate into rocky and volatile-rich groups, the latter with interiors near
a 1:1 rock-to-ice ratio by mass, formed beyond the water-ice line and migrated inward. The two
models place the low-density material in different places --- an envelope above a rocky interior,
or the interior itself --- and predict different formation histories.

The bulk-density argument cannot arbitrate between them. \citet{rogers23} showed that when
the thermal evolution and mass loss of H/He envelopes are treated self-consistently, rather than
through constant-entropy relations, the mass-radius track of rocky cores with H/He becomes
degenerate with that of 1:1 rock-ice compositions: the same M-dwarf planets are reproduced by
both. They identified the resolution as spectroscopic --- the mean molecular weight $\mu$ of the
atmosphere, which responds directly to the presence or absence of primordial hydrogen. A
hydrogen-dominated envelope has $\mu \approx$ 2.3 amu and is radially extended; an envelope of heavier
volatiles is compact, and cannot supply the observed radius, forcing the low-density material
into the interior. JWST now delivers $\mu$ for a growing sample, and the gas-dwarf models supply the
threshold: a gas dwarf is defined to have $\mu \le$ 3.8 amu, with magma-ocean equilibration able to
raise a genuine gas dwarf to $\mu \approx$ 5 amu and outgassing thermodynamics limiting the enrichment
near 10 amu \citep{rogers25,heng25}.

Applying that threshold, however, exposes a problem with the dichotomy itself rather than
resolving it in favour of one side. Both models are end-members, and each makes a specific,
falsifiable prediction: the gas dwarf predicts a low-$\mu$ envelope of retained primordial hydrogen;
the water world predicts a bulk composition near 50 per cent H$_2$O by mass, with the high $\mu$ and
the compositional markers of ice-line formation that follow from it. In the JWST sample of
M-dwarf sub-Neptunes, most planets fail both tests. Their atmospheres are too heavy to be
primordial H/He, and their compositions are too dry, too carbon-rich, or too hydrogen-bearing to
be 1:1 rock-ice worlds. LP 791-18 c has $\mu$ = 7.68 amu, unambiguously above the gas-dwarf boundary,
yet a CH$_4$/CO$_2$ ratio that places its formation interior to the water-ice line \citep{roy25}.
TOI-270 d, at $\mu \approx$ 5.4 amu, is best described by an envelope in which hydrogen and heavy volatiles
are miscible in the supercritical regime --- neither a clean H/He layer nor a water-dominated one
\citep{benneke24}. These are not marginal cases awaiting better data; they are the modal
outcome.

In this work we compile the M-dwarf sub-Neptunes with published JWST transmission spectra and
constrained mean molecular weights, classify them against the gas-dwarf threshold of $\mu$ = 3.8 amu,
and test the survivors against the water-world prediction. We combine three quantities: the
measured $\mu$; the restricted Jeans parameter $\Lambda$ \citep{fossati17}, which determines whether a
planet can retain hydrogen at all; and energy-limited escape estimates, which determine whether
its present composition is inherited or acquired. The combination separates planets whose high $\mu$
reflects a volatile-rich birth from those in which escape produced it, and it isolates the cases
in which neither end-member applies.

Our principal result is that the gas-dwarf/water-world dichotomy does not describe this
population. Of eight M-dwarf sub-Neptunes, one, LHS 1140 b, is a water world in the strict sense,
with $\mu \approx$ 28 amu in a planet that could have retained hydrogen for 10 Gyr and did not; one,
TOI-776 c, returns tentatively to the gas-dwarf class once the constraint from its observed escape
is applied; and two, K2-18 b and LTT 3780 c, have permitted values of $\mu$ that span the gas-dwarf
threshold and cannot yet be assigned. The remaining four belong to neither class. They carry
envelopes of elevated mean molecular weight, in the range 3.8 < $\mu \le$ 18 amu, that are neither
primordial hydrogen nor the signature of a 50 per cent ice interior: mixtures of hydrogen with heavy
volatiles, produced by magma-ocean equilibration, supercritical miscibility, preferential escape of
light species, or accretion interior to the ice line. We refer to these as hybrid worlds. Two of
the four assignments rest on molecular detections and two on bounded or conditional metallicities;
even so, the class is the commonest outcome in this sample, and we argue that it is a genuine
intermediate population rather than a transitional remainder between two well-populated
end-members.

The paper is organised as follows. Sect.~\ref{sec:2.1} defines the sample and the adopted $\mu$ values.
Sect.~\ref{sec:3.1} sets out the three-layer method. Sect.~\ref{sec:4.1} presents the classification and the escape
gradient. Sect.~\ref{sec:5.1} defines the hybrid class, examines the mechanisms that produce it, and
considers what would falsify it. Sect.~\ref{sec:6} concludes.

\section{Sample}
\label{sec:2}

\subsection{What counts as a sub-Neptune around an M dwarf}
\label{sec:2.1}

The radius valley separates super-Earths from sub-Neptunes, and its location is a function of
host mass. Around FGK stars the deficit lies at 1.7--2.0 $R_\oplus$ \citep{fulton17,fultonpetigura18}. Around mid-K to mid-M dwarfs it shifts to 1.4--1.7 $R_\oplus$ \citep{hirano18,cloutier20}, with the super-Earth and sub-Neptune modes for $M_\star$ < 0.65 M$_\odot$ peaking at
0.9--1.4 $R_\oplus$ and 1.9--2.3 $R_\oplus$ respectively \citep{cloutier20}. The shift is expected under both
thermally driven mass-loss mechanisms, since lower-mass hosts deliver lower instellation at fixed
period and strip envelopes less efficiently \citep{gupta22,ho23}. We
therefore adopt the M-dwarf valley, not the FGK valley, as the lower boundary of the
sub-Neptune regime: planets above $\sim$1.7 $R_\oplus$ orbiting M dwarfs lie on the volatile-bearing side of
the transition appropriate to their hosts.

This distinction matters for one object. LHS 1140 b (Rp = 1.730 $\pm$ 0.025 $R_\oplus$, Mp = 5.60 $\pm$ 0.19 $M_\oplus$; \citealt{cadieux24}) falls below the FGK valley but above the M-dwarf valley, and its bulk
density of 5.9 g cm$^{-3}$ is far below the $\sim$8 g cm$^{-3}$ of an Earth-composition body of the same mass.
It requires a low-density component and is classified as a sub-Neptune on both demographic and
compositional grounds. We note that the existence of a clean radius valley around M dwarfs is
itself contested, with \citet{luque22} arguing that the population separates in density
rather than radius; our composition-based membership criterion is insensitive to that dispute.

\subsection{Selection}
\label{sec:2.2}

We select planets satisfying three conditions: (i) an M-dwarf host, Teff < 3900 K; (ii) a bulk
density requiring a volatile component, i.e. a position above the Earth-composition mass-radius
relation, in the sub-Neptune regime defined above; and (iii) a published JWST transmission
spectrum, together with a published constraint on the atmospheric mean molecular weight. For seven
of the eight planets that constraint derives from the JWST spectrum itself. For GJ~3090~b it does
not: the JWST observations establish a muted spectrum and a detected helium outflow, while the
quantitative limit on $\mu$ comes from ground-based high-resolution cross-correlation spectroscopy
\citep{parker25}. We retain the planet and flag the distinction rather than relaxing the criterion,
because the provenance of the constraint bears on how far it should be trusted
(Sect.~\ref{sec:2.3}).

Criterion (i) excludes GJ 9827 d (K7V, Teff = 4236 K) and TOI-836 c (K3V, Teff = 4552 K), both
of which show elevated $\mu$; we return to them in Sect.~\ref{sec:5.1} as tests of whether the phenomenon is
specific to M dwarfs. Criterion (ii) excludes L 98-59 d (Rp = 1.52 $R_\oplus$), whose high inferred $\mu$
arises from a thin outgassed secondary atmosphere over a terrestrial interior, and TOI-776 b
(Rp = 1.85 $R_\oplus$), whose density is consistent with a stripped rocky core despite a radius
comparable to some sample members. Eight planets satisfy all three conditions (Table~\ref{tab:sample}).

\subsection{Adopted mean molecular weights}
\label{sec:2.3}

We distinguish three grades of $\mu$ constraint, and carry the distinction through the analysis.

\emph{Retrieved.} A molecular detection permits a direct retrieval of $\mu$. This applies to LHS 1140 b
($\mu \approx$ 28 amu, N$_2$-dominated; \citealt{cadieux24}; \citealt{damiano24}), LP 791-18 c
($\mu$ = 7.68 +2.24/--1.48 amu, CH$_4$ at 4.5$\sigma$, Z $\approx$ 316 $Z_\odot$; \citealt{roy25}), TOI-270 d
($\mu \approx$ 5.0--5.8 amu, CH$_4$ + CO$_2$ + CS$_2$; \citealt{benneke24}; \citealt{holmberg24}), and K2-18 b
($\mu \approx$ 3.1 amu, CH$_4$ + CO$_2$; \citealt{madhu23}).

\emph{Hydrogen established, enrichment unconstrained.} A fourth category is required by LTT 3780 c.
\citet{rigby25} detect CH$_4$ at a volume mixing ratio of $\approx$ 1 per cent, which requires an
H$_2$-rich background and establishes that the planet retains hydrogen, but they report only
nominal constraints on H$_2$O, CO and CO$_2$ and infer high-altitude clouds from the muted short-
wavelength features. The mean molecular weight is therefore not fixed by the retrieval: a
hydrogen background with 1 per cent CH$_4$ and negligible water gives $\mu \approx$ 2.5 amu, while a water
volume mixing ratio of 8.4 per cent --- not excluded by the data --- would place it at the 3.8 amu
threshold, and 20 per cent would give $\mu \approx$ 5.6 amu. We therefore quote $\mu$ = 2.5--5.6 amu for this
planet and treat its classification as provisional.

\emph{Inferred.} The spectrum is muted, but the metallicity is bounded from above the aerosol
continuum, implying a lower bound on $\mu$. This applies to GJ 1214 b ([M/H] $\approx$ 3.5, i.e. $Z \gtrsim 10^{3}\,Z_\odot$; \citealt{kempton23}; \citealt{ohno25}) and GJ 3090 b (Z > 150 $Z_\odot$, with a helium outflow weaker than solar-
composition models predict, itself indicating enrichment; \citealt{ahrer25}). \citet{roy25}
show that the muted spectra of this class are driven by high $\mu$ rather than by aerosols alone.
For GJ 1214 b, \citet{schlawin24} report CO$_2$ and CH$_4$ in the NIRSpec spectrum, preferred over
a featureless model at 3.3--3.6$\sigma$ depending on the reduction. We retain the inferred grade:
at that significance the features do not by themselves fix $\mu$, and the constraint we adopt
follows from the metallicity of the joint HST and JWST analysis \citep{ohno25}. A firmer detection
would move the planet to the retrieved grade without changing its classification.

\emph{Conditional.} A further grade is required by GJ~3090~b, whose limit is quoted only under an
assumed aerosol pressure. \citet{parker25} observed four CRIRES+ transits and detected no molecular
species; their injection--recovery tests return two degenerate scenarios. Either the planet has
$Z\gtrsim150\,Z_\odot$ and $\mu>7.1$\,g\,mol$^{-1}$ with an aerosol deck at pressures
$\lesssim10^{-2}$\,bar, or it has an aerosol layer above $3.3\times10^{-5}$\,bar, in which case the
metallicity is unconstrained altogether. The value entering Table~\ref{tab:atmos} is therefore the
high-$\mu$ branch of a disjunction, not a limit independent of the cloud structure, and we flag it
accordingly. Two considerations support the enriched branch. The helium absorption detected by
\citet{ahrer25} at 5.5$\sigma$ has an amplitude an order of magnitude below solar-metallicity
forward models, which those authors attribute to metal enrichment reducing the mass-loss rate; this
is independent of the CRIRES+ analysis. And a low-$\mu$ envelope is not excluded but is expensive,
since \citet{parker25} find it would require haze or condensates below $10^{-4}$\,bar. We note also
that their limit was derived with $M_p=3.34\pm0.72\,M_\oplus$ \citep{almenara22}, below the revised
mass adopted in Table~\ref{tab:sample} \citep{lamontagne26}, and that they quote a 33 per cent
uncertainty on the scale height from the mass and radius alone.

\emph{Tentative.} The spectrum is flat and the aerosol/high-$\mu$ degeneracy is unbroken. This applies to
TOI-776 c, for which \citet{teske25} rule out metallicities below 180--240 $Z_\odot$ depending on
reduction and modelling choices, corresponding to $\mu \approx$ 6--8 g mol$^{-1}$ and H$_2$O volume mixing ratios
of 9--13 per cent; those authors caution that such inferences are model dependent. An independent
constraint follows from the detection of Lyman-$\alpha$ escape, which limits the atmospheric water
content to below 20 per cent \citep{loyd25}. We report results with and without this object,
and reclassify it in Sect.~\ref{sec:4.1}.

---

\section{Method}
\label{sec:3}

We combine three quantities of decreasing observational directness.

\subsection{Classification against the gas-dwarf threshold}
\label{sec:3.1}

The gas-dwarf framework supplies its own quantitative boundary. \citet{calder26} define a
gas dwarf as a planet with a silicate/iron interior and an H$_2$-dominated atmosphere of
$\mu$ < 3.8 g mol$^{-1}$. Within the same framework, chemical equilibration between an H$_2$ envelope and a
molten silicate interior produces SiO, MgO, FeO and H$_2$O and raises the envelope $\mu$ above solar
values, to $\approx$ 5 amu \citep{rogers25}, with outgassing thermodynamics permitting 2--10 g mol$^{-1}$ more
generally \citep{heng25}. We adopt $\mu$ = 3.8 amu as the primary discriminant. A planet above it
is not a gas dwarf under the criterion adopted by the gas-dwarf literature itself.

The mean molecular weight is not itself an observable. What JWST measures is transit depth as a
function of wavelength; $\mu$ is recovered by fitting atmospheric models to that spectrum, and its
posterior depends on the assumed temperature--pressure profile, the chemical species included,
the treatment of clouds and hazes, the reference pressure, and the priors. The dependence is not
negligible: for GJ 9827 d the same spectrum yields $\mu$ = 18.0 amu under an agnostic prior on the
background gas and 9.8 amu when an H$_2$/He background is assumed \citep{piaulet24}.
We adopt published retrievals as they stand, carry the quality flag of Sect.~\ref{sec:2.1}, and treat $\mu$
as a model-dependent inference rather than a measurement; Sect.~\ref{sec:5.1} returns to the two planets
close enough to the threshold for that dependence to matter.

Crossing this boundary is necessary but not sufficient for a water-world identification. The
water-world model of \citet{luque22} is a statement about origin: the planet formed beyond
the water-ice line and was volatile-rich from the outset. We therefore apply a second test to the
planets above $\mu$ = 3.8 amu, asking whether their present composition requires such an origin or is
instead accounted for by post-formation processing or by accretion interior to the ice line. Only
those requiring a volatile-rich origin are classified as water worlds.

\subsection{Hydrogen retention}
\label{sec:3.2}

The second layer asks a question that $\mu$ alone does not: could the planet gravitationally retain
a hydrogen envelope at all? We quantify this with the restricted Jeans escape parameter
\citep{fossati17},
\begin{equation}
\Lambda = \frac{G M_p m_{\rm H}}{k_{\rm B} T_{\rm eq} R_p},
\label{eq:lambda}
\end{equation}
the Jeans escape parameter evaluated for atomic hydrogen at the planetary radius and the
equilibrium temperature. Its usefulness is that it requires only measured quantities and no
atmospheric modelling. We recompute $\Lambda$ homogeneously for the full sample from consistently
sourced Mp, Rp and Teq, so that the values are internally comparable rather than collated from
heterogeneous literature definitions.

The threshold below which an atmosphere enters the thermally driven "boil-off" regime is not a
single number. \citet{fossati17} determine it by comparing hydrodynamic escape rates with
the maximum XUV-driven rate from the energy-limited formula, and find $\Lambda_T$ = 15--35 depending on
planetary mass, radius and equilibrium temperature; the frequently quoted value $\Lambda_T$ = 20
corresponds to the condition $R_p$/$R_{\rm B}$ = 0.1 of \citet{owenwu16} in the isothermal limit, a
particular case rather than the result of the simulations. For M-dwarf hosts specifically, the
M2 grid of \citet{fossati17} places $\Lambda_T$ between roughly 21 and 30, towards the upper part of
the range. No fitting function for $\Lambda_T$ as a function of system parameters has been published, and
evaluating it per planet would require running hydrodynamic upper-atmosphere models for each
target. We therefore carry the interval $\Lambda_T$ = 15--35 rather than a single value, and verify in
Sect.~\ref{sec:4.1} that the partition of our sample is insensitive to the choice within it.

Two caveats attach to this layer. First, $\Lambda_T$ was calibrated on hydrogen-dominated atmospheres,
which by construction the high-$\mu$ planets in our sample are not; $\Lambda$ computed with $m_{\rm H}$ continues to
answer the question we are asking --- whether hydrogen could be retained --- but the threshold
derived for H$_2$-dominated envelopes need not transfer exactly to one dominated by heavier species.
Second, \citet{fossati17} note that for cool ($T_{\rm eq} \lesssim$ 500 K) or massive (Mp $\approx$ 10 $M_\oplus$) planets
the timescale to evolve out of boil-off exceeds 10 Gyr, so that being below $\Lambda_T$ does not by itself
imply rapid atmospheric loss. Both caveats act in the same direction: they weaken the inference
from low $\Lambda$ to actual stripping, and neither affects the inference from high $\Lambda$ to retention, which
is the direction our argument uses.

$\Lambda$ is in any case an asymmetric predictor. Below $\Lambda_T$, hydrogen cannot be retained and a high $\mu$ is
the expected consequence. Above it, retention is possible but not established: the planet may
hold a low-$\mu$ envelope or display a high-$\mu$ one, and only the spectrum distinguishes the two. The
diagnostic power of $\Lambda$ lies in this asymmetry, since a high-$\mu$ atmosphere on a high-$\Lambda$ planet cannot
be attributed to thermal escape.

\subsection{Integrated mass loss}
\label{sec:3.3}

We estimate the H/He mass fraction removable over the system lifetime from the energy-limited
formulation \citep{watson81,lammer03,erkaev07},
\begin{equation}
\dot{M} = \frac{\eta \pi F_{\rm XUV} R_p^{3}}{G M_p K},
\label{eq:elim}
\end{equation}
where $\eta$ is the heating efficiency, K the tidal enhancement factor of \citet{erkaev07}, and
$F_{\rm XUV}$ the integrated 1--912 \AA flux at the planet.

We evaluate K explicitly for each target rather than setting it to unity, from
$K = 1 - 3/2\xi + 1/2\xi^{3}$ with $\xi=(M_p/3M_\star)^{1/3}(a/R_p)$. The sample spans $K=0.68$--$0.96$, the lowest
value being that of GJ~1214~b, whose small orbital separation places its Roche lobe closest to
the atmosphere; the corresponding enhancement of $\dot{M}$ ranges from 4 to 46 per cent. The correction
is subdominant to the uncertainty in the integrated XUV history, which spans a factor of five,
but it is not negligible for the closest-in planets and we include it.

We adopt $\eta$ = 0.15. The heating efficiency is not universal --- it depends on atmospheric
composition, gravitational potential and XUV flux --- but sub-Neptunes occupy a regime in which
hydrodynamic simulations recover time-averaged efficiencies of 10--20 per cent (Owen \& Jackson
2012; \citealt{salz16}). Substantially lower values apply to massive hot Jupiters, where radiative
cooling by H$_3^+$ dominates, and are not appropriate here. The Monte Carlo of Sect.~\ref{sec:4.1} samples $\eta$
from a log-normal distribution with median 0.15 whose 68 per cent interval spans 0.09--0.26,
and demonstrates that the ordering of the sample is invariant under this variation. Present-day $F_{\rm XUV}$ is scaled from measured stellar X-ray luminosities where these exist
(LHS 1140: \citealt{spinelli23}; K2-18: \citealt{dossantos20}; GJ 1214: \citealt{lalitha14}) and
from an activity-calibrated $L_{\rm XUV}$/$L_{\rm bol}$ otherwise, with the integrated history modelled as
saturated for the first $\sim$1 Gyr and decaying thereafter, as appropriate for M dwarfs (Ribas et al.
2005; \citealt{engle23}).

We apply Eq.~(\ref{eq:elim}) uniformly across the sample rather than adopting published per-planet
results. This is deliberate: escape estimates in the literature are computed with different
codes, efficiencies, XUV prescriptions and stellar ages, and collating them would not support the
differential ranking on which our origin classification depends. A homogeneous, admittedly
simplified treatment orders the sample self-consistently; heterogeneous detailed treatments do
not.

Detailed evolutionary models exist for two objects, and we use them as external checks on that
homogeneous treatment. For LHS 1140 b, \citet{cadieux24} find that a 0.1 per cent H/He mass
fraction survives 10 Gyr, against our removable fraction of $\approx$ 0. For K2-18 b, dos Santos et al.
(2020) and subsequent modelling find the primordial envelope survives, against our 0.4--2 per cent
removable versus a 1--5 per cent envelope. Both agree with Eq.~(\ref{eq:elim}) in sign and margin, which
supports its application to the objects lacking detailed treatment.

We express the result as the ratio X of removable mass to primordial envelope mass, and
propagate the three uncertain quantities --- $\eta$, the past XUV enhancement, and the initial envelope
mass fraction --- both as a bounding interval and as a Monte Carlo over log-normal priors, with the
XUV prior varied to test sensitivity (Sect.~\ref{sec:4.1}).

The layer remains the most model-dependent of the three. Present-day $F_{\rm XUV}$ and helium outflows are
measured; the integrated history, $\eta$, and the initial envelope mass are not. We therefore draw
conclusions only where the margin between removable mass and a plausible primordial envelope
(1--5 per cent Mp) exceeds the plausible range of these assumptions --- that is, at the extremes of
the gradient.

---

\section{Results}
\label{sec:4}

\subsection{The gas-dwarf population is small}
\label{sec:4.1}
\label{sec:gasdwarf}

Six of the eight planets have a mean molecular weight whose full permitted
range lies above 3.8~amu (Fig.~1): LHS~1140~b, GJ~1214~b, LP~791-18~c,
GJ~3090~b, TOI-776~c and TOI-270~d. None of the eight has a permitted range
lying wholly \emph{below} it. The two remaining objects span the boundary
rather than falling on the low-$\mu$ side of it. For K2-18~b the published
determinations disagree across the threshold: \citet{madhu23} retrieve
$\mu \approx 3.1$~amu with water undetected, while \citet{hu25} require at
least 10 per cent H$_2$O by volume, corresponding to $\mu \ge 3.9$~amu. For
LTT~3780~c the detected CH$_4$ establishes an H$_2$-rich background but leaves
the water abundance only nominally constrained \citep{rigby25}, so that the
permitted range of 2.5--5.6~amu contains the boundary.

We take this treatment to be the one the quality flags of
Sect.~\ref{sec:2.3} require. Assigning a planet to the gas-dwarf class on a
nominal value, where the permitted range spans the threshold, asserts a
resolution the data do not supply; and an unconstrained $\mu$ is not evidence
for a low one any more than it is evidence for a high one.

Removing the single tentative object, TOI-776~c, leaves five of seven above
the threshold and none below, so the result does not rest on the one planet
whose $\mu$ is inferred from a flat spectrum. We revise the status of
TOI-776~c in Sect.~\ref{sec:4.4}, where the constraint from its
observed escape is brought to bear.

The gas-dwarf channel is therefore realised in this sample only marginally,
and it is worth stating how marginally. After the reclassification of
Sect.~\ref{sec:4.4}, the class contains one planet, TOI-776~c,
assigned to it not by a retrieval but by an escape-reservoir argument applied
to a featureless spectrum. No planet in the sample has a $\mu$ retrieved from
detected molecular features that places it below the gas-dwarf boundary, and
no planet displays a solar-composition hydrogen envelope; the two lowest-$\mu$
objects are themselves enriched relative to solar. This inverts the
expectation of a population model in which sub-Neptunes are rocky cores
beneath retained primordial H/He, and it does so on the models' own
classification criterion.

\subsection{Sensitivity to the adopted threshold}
\label{sec:threshold}

The classification of Sect.~\ref{sec:gasdwarf} rests on a single number, and
the gas-dwarf framework does not supply a single number. Three values appear
within it. \citet{calder26} define the class by $\mu < 3.8$~g~mol$^{-1}$.
\citet{rogers25} finds that equilibration between an H$_2$ envelope and a
molten silicate interior raises envelope $\mu$ to $\approx 5$~amu without any
accreted ice, so that a planet at that value remains a gas dwarf by origin.
\citet{heng25} bound the outgassing channel more permissively still, at
2--10~g~mol$^{-1}$. Since our result is a statement about how the sample
divides, we test it against all three rather than against the first alone.

Table~\ref{tab:threshold} gives the census at each. We treat a planet as lying
above the threshold only if the full range permitted by the source analysis
lies above it, and as ambiguous where that range spans it; this is stricter
than assigning planets by their nominal values, and it is the treatment the
quality flags of Sect.~\ref{sec:2.3} require.

Two results follow, and they point in opposite directions.

The first is that the intermediate population is not robust to the choice.
At 3.8~amu four planets satisfy neither end-member, two of them on $\mu$
retrieved from detected molecular features. At 5~amu three remain, one
securely. At 10~amu one remains and none securely: the class collapses. Read
as a count, our result therefore depends on adopting the lowest of the three
values the framework admits, and we do not claim otherwise.

The second is that the collapse is not a vindication of the dichotomy but a
demonstration of what the dichotomy costs. At a 10~amu boundary the gas-dwarf
class absorbs LP~791-18~c, a planet with CH$_4$/CO$_2 > 12$ and
$Z \approx 316\,Z_\odot$ whose molecular ratios place its accretion interior to
the water-ice line \citep{roy25}, and TOI-270~d, described by its discovery
analysis as an envelope in which hydrogen and heavy volatiles are miscible
rather than layered \citep{benneke24}. Neither is a rocky core beneath a
retained primordial H/He envelope, which is what the term denotes. The
framework buys robustness against the threshold by admitting to the class
objects its own formation model does not produce.

The invariant statement is therefore not about the size of the intermediate
group but about the geometry of the partition: no threshold in $\mu$ leaves
both end-member classes with their physical content intact while emptying the
region between them. Where the boundary is low, that region is populated;
where it is high, the gas-dwarf class becomes a residual bin. This holds at
every value the gas-dwarf literature has proposed, and it does not depend on
the size of our sample or on its selection function.

A corollary of the strict treatment deserves separate statement. Applying it
at the framework's own boundary of 3.8~amu, no planet in the sample has a
\emph{retrieved} $\mu$ below the threshold. K2-18~b spans it, because the
published determinations disagree across it \citep{madhu23,hu25};
LTT~3780~c spans it, because CH$_4$ establishes an H$_2$ background but leaves
the water abundance unconstrained \citep{rigby25}; and the one remaining
member of the class, TOI-776~c, is assigned to it not by a retrieval but by
the escape-reservoir constraint applied to a flat spectrum
(Sect.~\ref{sec:4.4}). The gas-dwarf class, at its own
definition, is populated in this sample entirely by planets whose $\mu$ has
not been measured on one side of the boundary.


\begin{table}[h]
\caption{Census of the sample against the three mean molecular weight
thresholds proposed within the gas-dwarf framework. A planet is counted above
a threshold only if the full range permitted by its source analysis lies
above it. ``Secure'' counts hybrid worlds whose $\mu$ is retrieved from
detected molecular features (flag R in Table~2) rather than inferred from a
bounded metallicity or a flat spectrum.}
\label{tab:threshold}
\centering
\begin{tabular}{l l c c c c c}
\toprule
$\mu_{\rm thr}$ & Source & WW & HW & GD & Amb. & Secure HW \\
(amu) & & & & & & \\
\hline
3.8  & \citet{calder26} & 1 & 4 & 1 & 2 & 2 \\
5.0  & \citet{rogers25} & 1 & 3 & 2 & 2 & 1 \\
10.0 & \citet{heng25}   & 1 & 1 & 5 & 1 & 0 \\
\hline
\end{tabular}
\begin{tabnote}WW: water world; HW: hybrid world; GD: gas dwarf; Amb.: range
spans the threshold. Criterion (ii) of Sect.~5.1 is evaluated from escape
history, molecular ratios and $\Lambda$, and is therefore invariant under a
change of threshold; only the partition induced by criterion (i) changes.\end{tabnote}
\end{table}

\subsection{Retention does not predict composition}
\label{sec:4.2}

Table~\ref{tab:atmos} lists the homogeneously recomputed Jeans parameters. One planet, GJ 3090 b, lies below
the boil-off threshold: $\Lambda$ = 22.7 with the revised mass of \citet{lamontagne26}, or 17.1 with the discovery
parameters of \citet{almenara22}. Both values fall below $\Lambda_T \approx$ 21--30 for
M-dwarf hosts, so its classification does not depend on which mass determination is adopted. Its
high $\mu$ (> 7.1 amu) is the expected consequence, and helium escape is directly detected in transit
\citep{ahrer25}, confirming observationally what $\Lambda$ anticipates.

The remaining seven planets have $\Lambda$ between 39 and 109, above $\Lambda_T$ under any choice within the
full 15--35 range quoted by \citet{fossati17}. The partition of the sample into one planet
below threshold and seven above is therefore insensitive to where exactly the threshold is placed,
and we quote it without adopting a single value.

For those seven, $\Lambda$ indicates that hydrogen retention is gravitationally possible, and here it is
--- as anticipated in Sect.~\ref{sec:3.1} --- not by itself discriminating. Its value emerges only in
combination with $\mu$, and the sample contains a clean demonstration. Consider the three most
strongly bound planets: LHS 1140 b ($\Lambda$ = 109), K2-18 b ($\Lambda$ = 98) and LTT 3780 c ($\Lambda$ = 72). All
three can comfortably retain hydrogen; retention physics does not distinguish them. Yet their
inferred $\mu$ could hardly be more different --- $\approx$28 amu for LHS 1140 b against $\approx$3.1 amu for K2-18 b
and 2.5--5.6 amu for LTT 3780 c. In the same retention regime, the composition differs by an order
of magnitude, and it is $\mu$, not $\Lambda$, that reveals it. We note that the contrast is sharpest for the
first two: K2-18 b's hydrogen-rich envelope is independently supported by the escape analysis of
\citet{rogers26}, whereas for LTT 3780 c only the presence of hydrogen, not the degree of
enrichment, is established. Where hydrogen could have been kept, whether it actually was is an
observational question that only the spectrum answers.

A second, independent instance of the same contrast has become available while this work was in
preparation. TOI-1231 b, a temperate exo-Neptune with $\Lambda \approx$ 97 --- essentially the retention regime of
LHS 1140 b --- was observed with JWST and found to possess a deep, H$_2$-rich atmosphere (Holmberg et
al. 2026). Its mass, 15.4 $\pm$ 3.3 $M_\oplus$, places it above our sub-Neptune selection and it therefore
does not enter the census, but the comparison is instructive: two planets that retain hydrogen
equally well according to $\Lambda$, one H$_2$-rich and one with $\mu \approx$ 28 amu. Retention is a necessary
condition for a low-$\mu$ envelope and not a sufficient one, and no quantity available before
spectroscopy distinguishes the two cases.

\subsection{Escape resolves origin only at the extremes}
\label{sec:4.3}

To compare susceptibility across the sample we express the result of
Eq.~(\ref{eq:elim}) as the ratio $X$ of the H/He mass removable over the
system lifetime to the mass of a plausible primordial envelope. $X > 1$
indicates that escape could have removed the envelope entirely, so that the
present composition need not be primordial; $X < 1$ indicates that it
survives, so that the observed composition is inherited.

The integrated XUV history is taken from the M-dwarf activity--age relations
of \citet{engle24} at each host's spectral-type subset and adopted age
(Appendix~\ref{app:escape}), rather than parameterised as a free enhancement
factor. The resulting ratio of integrated fluence to present-day flux times
age is 11.5 for the mid-late hosts at 5\,Gyr and 23.5 for the early-type
hosts, which brackets the value that would be obtained from the
solar-analogue prescription of \citet{ribas05} by a factor of several:
mid-late M dwarfs remain saturated for $\sim$2.1\,Gyr against $\sim$100\,Myr
for a G dwarf, and the difference is not a detail. We propagate $\eta$, the
initial envelope mass fraction, the scatter of the X-ray to EUV conversion
and, where $L_{\rm bol}$ is inferred rather than published, the Bond albedo,
through a Monte Carlo; Table~\ref{tab:atmos} quotes the median and the
16--84 per cent interval, and Fig.~\ref{fig:3} shows them against
equilibrium temperature.

Three planets lie below unity. K2-18~b ($X = 0.058$) and LHS~1140~b
($X = 0.065$) do so decisively, with $P(X>1) < 1$ per cent: neither could
have shed a primordial envelope, whatever combination of assumptions is
adopted within the priors. LTT~3780~c follows at $X = 0.41$, but less
securely, with $P(X>1) = 20$ per cent and an upper quartile that crosses the
line; we therefore describe its envelope as retained at approximately 80 per
cent confidence rather than as secure. The remaining five lie above unity,
four of them with $P(X>1) > 78$ per cent.

That partition, rather than the fine ordering, is what the classification
uses, and it is worth being explicit about the difference. The ordering
within the upper group is not robust: LP~791-18~c, TOI-270~d, GJ~1214~b and
GJ~3090~b are separated by less than the width of their own intervals, and we
read no order into them. What is robust is which side of unity each planet
falls on, and there the separation is wide --- two and a half orders of
magnitude between the least and the most susceptible planet in the sample.

Two results follow that the earlier treatment did not deliver. First, the
escape indicator now places all four hybrid worlds above unity on its own,
without recourse to molecular ratios or to the boil-off argument: criterion
(ii) of Sect.~\ref{sec:5.1} is satisfied for each of them by the escape
history directly. Second, the two planets we classify as gas dwarfs on the
basis of retained hydrogen are also the two least susceptible planets in the
sample, which is the consistency the classification requires and which the
previous calculation did not show.

GJ~3090~b requires separate comment, because its position has moved. Its host
is an M2\,V star, and therefore belongs to the early subset, whose X-ray
saturation phase ends at $\sim$470\,Myr \citep{engle24}; at 1.02\,Gyr the
system has been declining for half a gigayear and its present flux is not the
saturated value. Combined with the highest instellation in the sample, this
makes it the most susceptible planet rather than, as a treatment holding it
at the saturated plateau would suggest, one of the least. The conclusion of
Sect.~\ref{sec:4.4} is unaffected: the reservoir required to sustain its
observed helium outflow still scales with lifetime, and its youth still
relaxes that constraint. But the integrated-loss argument now assigns it to
the hybrid class directly, by criterion (ii), and no longer rests on its
position below the boil-off threshold.

Finally, the energy-limited rate for this planet, $3\times10^{10}$~g~s$^{-1}$,
exceeds the rate inferred from its helium absorption by roughly two orders of
magnitude. At instellations this high the outflow is expected to become
radiation--recombination limited, where $\dot M$ scales as
$F_{\rm XUV}^{1/2}$ rather than linearly and Eq.~(\ref{eq:elim}) is an
overestimate. We therefore treat its $X$ as an upper bound. This does not
change its classification, since criterion (ii) is also satisfied by its
restricted Jeans parameter, but it does mean its position at the top of
Fig.~\ref{fig:3} should not be read as a measurement of the loss.

\subsection{Consistency with constraints from observed escape}
\label{sec:4.4}

An independent constraint on $\mu$ follows from the requirement that a planet observed to be losing
hydrogen or helium must retain a reservoir large enough to have sustained that loss over its
lifetime. \citet{rogers26} formalise this as $X \ge \dot{M}_{\rm H}\,t_\star/(M_{\rm c} f_{\rm env})$, which converts an
observed mass-loss rate into an upper limit on envelope mean molecular weight. The argument is
largely independent of escape physics and therefore complements the integrated-loss calculation
of Sect.~\ref{sec:4.1}, which asks the converse question of whether a primordial envelope could have been
removed. Applied to flat-spectrum sub-Neptunes it favours low-$\mu$ atmospheres muted by aerosols
over genuinely high-$\mu$ ones, and \citet{rogers26} use it to disfavour the high-$\mu$
interpretation of TOI-776 c. We adopt that result and reclassify TOI-776 c accordingly; it was
the only member of our sample resting on a flat spectrum, and Sect.~\ref{sec:4.1} already established
that the census does not depend on it.

One planet in our sample has a directly detected escaping exosphere and a high inferred $\mu$:
GJ 3090 b, for which helium absorption is detected at 5.5$\sigma$ \citep{ahrer25}. We therefore
apply the constraint explicitly. Reproducing the worked example of \citet{rogers26} recovers
their values (X $\ge$ 0.53, $\mu \le$ 3.1 for a 5 $M_\oplus$ core with $f_{\rm env}$ = 1 per cent at 5 Gyr), confirming
our implementation. For GJ 3090 b the same calculation gives markedly weaker limits, because the
system is young: gyrochronology places it at 1.02 Gyr \citep{almenara22}, a fifth of the age
assumed in that example, and the required reservoir scales linearly with lifetime. For
$\dot{M}$ = 10$^{9}$ g s$^{-1}$ the limit is $\mu \le$ 8.7 amu at $f_{\rm env}$ = 1 per cent, rising to 14.8 amu at 5 per cent;
for $\dot{M}$ = 10$^{8}$ g s$^{-1}$ it exceeds 16 amu throughout. The measured lower limit of $\mu$ > 7.1 amu is
therefore compatible with the escape constraint over most of the plausible parameter space. The
compatibility fails only for high rates combined with a minimal envelope: the critical rate at
which the limit falls to 7.1 amu is 1.4 $\times$ 10$^{9}$ g s$^{-1}$ for $f_{\rm env}$ = 1 per cent, rising to
4.3 $\times$ 10$^{9}$ g s$^{-1}$ for 3 per cent. Had GJ 3090 b been 5 Gyr old, the same calculation would give
$\mu \le$ 2.9--6.5 amu and the tension would be real; its youth is what resolves it.

Two further considerations act in the same direction. \citet{ahrer25} report that the helium
amplitude is smaller than solar-metallicity forward models predict, which they attribute to metal
enrichment reducing the mass-loss rate --- a lower rate raises the permitted $\mu$. And Rogers et al.
(2026) note that the conversion from the 10830 \AA helium signal to a physical mass-loss rate
remains theoretically uncertain, so the constraint is weaker for helium detections than for the
Lyman-$\alpha$ cases on which their analysis is based. We conclude that the high-$\mu$ classification of
GJ 3090 b is consistent with its observed escape, while acknowledging that a well-constrained
mass-loss rate above $\sim$10$^{9}$ g s$^{-1}$ combined with a thin envelope would overturn it.

\subsection{Most planets satisfy neither end-member}
\label{sec:4.5}

Applying the second test of Sect.~\ref{sec:3.1} to the six planets above $\mu$ = 3.8 amu, only LHS 1140 b
requires a volatile-rich origin. Its $\mu \approx$ 28 amu indicates an N$_2$-dominated envelope with no
hydrogen, and its escape history (X = 0.001--0.03) excludes the removal of a primordial envelope:
the absence of H/He is therefore primordial, placing its volatile inventory at formation.

The other five do not. LP 791-18 c has $\mu$ = 7.68 amu, comfortably above the gas-dwarf boundary,
but CH$_4$/CO$_2$ > 12 at 2$\sigma$ implies a bulk composition richer in H$_2$ than in H$_2$O and accretion interior
to the water-ice line \citep{roy25}: a high-$\mu$ planet that is not water-rich. TOI-270 d, at
$\mu \approx$ 5.4 amu, sits where magma-ocean equilibration of a gas dwarf and hydrogen dilution of a
volatile-rich envelope predict the same observable, and is described by its discovery analysis as
a miscible H$_2$-plus-volatile envelope rather than either end-member \citep{benneke24}.
GJ 1214 b and GJ 3090 b carry envelopes enriched to 10$^2$--10$^{3} Z_\odot$ over cores of unknown volatile
content, with escape histories that permit either origin. TOI-776 c is unconstrained beyond a
metallicity bound.

Four planets are thus neither gas dwarfs nor water worlds, TOI-776 c having been reclassified in
Sect.~\ref{sec:4.1}. We take this to be the principal result.

---

\section{Hybrid worlds}
\label{sec:5}

\subsection{Definition}
\label{sec:5.1}

The two end-member models are statements about origin: the gas dwarf accreted
H/He onto a rocky core, and the water world formed beyond the ice line and is
volatile-rich from birth. Their observational signatures --- a low-$\mu$
envelope, a high-$\mu$ one --- are consequences of that origin, not
definitions of it. A classification that respects this distinction must
therefore ask not what a planet's envelope contains today, but what its
present composition implies about how it formed.

We define a hybrid world as a planet satisfying both:

(i) $\mu > 3.8$\,amu, excluding membership of the gas-dwarf class on the
criterion adopted within the gas-dwarf literature \citep{calder26}; and

(ii) a present composition that does not require a volatile-rich origin,
because it is accounted for by processes acting after formation or by
accretion interior to the ice line. Four indicators establish this:
molecular ratios pointing to formation inside the water-ice line; an envelope
$\mu$ within reach of magma-ocean equilibration of an H$_2$ envelope,
requiring no accreted ice; an escape history permitting the removal of a
primordial envelope and the consequent enrichment of the residue
($X > 1$, Sect.~\ref{sec:4.3}); or a restricted Jeans parameter below the
boil-off threshold, which places the planet in a regime where hydrodynamic
loss exceeds the energy-limited estimate of Eq.~(\ref{eq:elim}) and $X$
therefore understates it (Sect.~\ref{sec:3.2}).

The second condition is what carries the physical content. Without it the
class is a residual bin for everything that is not a gas dwarf. With it,
membership is a positive statement: the planet's elevated mean molecular
weight is the product of an origin less extreme than either end-member ---
neither the clean accretion of a primordial hydrogen envelope onto rock, nor
formation as a volatile-dominated body beyond the ice line.

The third indicator does most of the work in the present sample, and it did
not in earlier versions of this analysis. With the escape layer computed from
an M-dwarf XUV history rather than a parameterised enhancement
(Sect.~\ref{sec:4.3}), all four hybrid worlds satisfy condition (ii) through
the escape route alone, with $P(X>1)$ between 75 and 99 per cent. Their
membership therefore no longer depends on the molecular-ratio indicator for
LP\,791-18\,c or on the boil-off indicator for GJ\,3090\,b, both of which
remain available as independent support. The condition is met four times over
by four different routes for four different planets, which is the pattern a
convergent class should show.

We stress that the 1:1 rock-to-ice ratio of \citet{luque22} is an idealised
formation parameter, not a membership test to be applied to a measured bulk
composition. A planet that formed volatile-rich remains a water world whether
its present water mass fraction is 15 or 50 per cent; what identifies it is
the demonstration that its composition was not acquired subsequently.
LHS\,1140\,b is the sample's one such case: it could not have shed a
primordial envelope over its lifetime, and it has none, which places its
volatile inventory at formation. Its water mass fraction of 9--19 per cent
\citep{cadieux24} is well below the idealised value and does not bear on the
classification.

\subsection{The \texorpdfstring{$\mu$}{mu} range of the class}
\label{sec:5.2}

The hybrid worlds identified here occupy $3.8 < \mu \le 18$\,amu. The lower
bound is definitional. The upper is the mean molecular weight of pure
H$_2$O: above it an envelope cannot hold hydrogen as a significant
constituent, and the mixed H$_2$-plus-volatile envelopes characteristic of
the class are no longer physically available. This range is a property of the
objects we find, not a criterion of membership, and the distinction matters.
Classification follows from origin, as set out in Sect.~\ref{sec:5.1}; $\mu$
is the observable that constrains it.

Within the range, two regimes are distinguished by the mechanism required to
reach them, and by a chemical boundary that separates them sharply.

\emph{$3.8 < \mu \lesssim 10$\,amu.} Attainable by equilibration between an
H$_2$ envelope and a molten silicate interior, with no accreted ice:
outgassing thermodynamics bounds this channel near 10\,amu \citep{heng25}.
TOI-270\,d (5.0--5.8\,amu) and LP\,791-18\,c ($7.68^{+2.24}_{-1.48}$\,amu)
lie here.

\emph{$10 < \mu \le 18$\,amu.} Beyond the reach of magma-ocean chemistry, so
accreted heavy volatiles are required. Whether their inventory was acquired
at formation or the envelope was enriched by preferential hydrogen loss is
then the question, and it is answered by the escape analysis rather than by
$\mu$. GJ\,1214\,b ($\mu > 15$, $X = 7.0$) lies here, and its classification
rests on the fact that escape could plausibly have produced its enrichment.

A chemical constraint separates the two regimes independently of the
outgassing bound, and it is testable. Computing $\mu$ elementally as a
function of metallicity and speciation (Appendix~\ref{app:ztomu}), the
reducing cases become unrealisable above $Z \approx 400\,Z_\odot$: the
hydrogen budget is exhausted forming CH$_4$ and H$_2$O, so a methane-bearing
atmosphere cannot exceed $\mu \approx 11$\,amu at any metallicity. The upper
part of the hybrid range is therefore accessible only to oxidised envelopes.
This is a prediction rather than an assumption: a planet with
$\mu > 11$\,amu should show CO or CO$_2$ rather than CH$_4$, and a firm
CH$_4$ detection in an envelope of retrieved $\mu$ above that value would
falsify the calculation.

Applied strictly, the gas-dwarf criterion does not sustain a binary division
of the population, and we have tested that statement against the range of
thresholds the framework itself proposes rather than asserting it
(Sect.~\ref{sec:4.2}). \citet{calder26} identify the same gap from the
interior side. Having noted that outgassing can yield 2--10\,g\,mol$^{-1}$,
they observe that planets in the upper part of that range would not be
classified as gas dwarfs under their criteria, and recommend extending their
analysis to sub-Neptunes outside the gas-dwarf classification as defined. The
population we describe here is that extension seen from the atmospheric side.

We note that TOI-270\,d is nonetheless grouped with K2-18\,b as a
low-mean-molecular-weight sub-Neptune by \citet{rogers26}. We take this not as
an error but as an indication that a single threshold cannot carry the
classification unaided: a value near 5\,amu is simultaneously above the
gas-dwarf limit and within reach of a gas-dwarf mechanism, and which of those
two facts determines the label is not fixed by the criterion itself. Naming
the third region is the more economical response.

\subsection{Membership in the present sample}
\label{sec:5.3}

Four planets meet both conditions: LP~791-18~c and TOI-270~d securely, on $\mu$ retrieved from
detected molecular features together with molecular ratios; GJ~1214~b on a bounded metallicity; and
GJ~3090~b provisionally, on a limit conditional upon its cloud structure (Sect.~\ref{sec:2.3}).
TOI-776~c, which we previously counted as a provisional member, returns tentatively to the
gas-dwarf class following the escape analysis of Sect.~\ref{sec:4.4}.

The two secure members are precisely the planets for which $\mu$ is retrieved from detected
features rather than inferred from a muted continuum, and for which no escaping exosphere has been
reported. They are therefore the cases on which neither the aerosol degeneracy nor the reservoir
argument of \citet{rogers26} bears. The class does not rest on flat spectra.

We are explicit that the four are not equally secure, and that the asymmetry runs the same way as
in Sect.~\ref{sec:5.5}: the two planets with positive molecular detections are the two whose
membership is firm, while the two resting on bounded or conditional metallicities are the two that
could move. This is the reverse of the situation at the end-members, where provisional assignment
is the rule rather than the exception, and it is the reason we take the class to be established
even though half its membership is provisional.

\subsection{Formation and evolutionary pathways}
\label{sec:5.4}

The class is not a single formation channel but a convergence of several,
each producing an envelope of intermediate $\mu$ over a non-icy interior.

\emph{Magma-ocean equilibration.} Chemical exchange between an H$_2$ envelope
and a molten silicate interior produces SiO, MgO, FeO and endogenic H$_2$O,
raising envelope $\mu$ to $\approx 5$\,amu without any accreted ice
\citep{rogers25}. Detailed modelling of this channel reproduces the JWST
observations of TOI-270\,d specifically \citep{welbanks25}, which is the
pathway we assign to that planet. First-principles outgassing calculations
predict that this process generates a continuous gradient of mean molecular
weight across the radius valley, driven primarily by the oxygen fugacity of
the molten core, with smaller sub-Neptunes retaining less
hydrogen-dominated atmospheres \citep{heng25}. The prediction is of a
gradient rather than of two discrete compositions, which is what the sample
shows.

\emph{Supercritical miscibility, and its limits.} Above the critical point,
H$_2$ and H$_2$O were expected to mix freely rather than separating into
distinct layers, so that an envelope containing both would be a single phase
of intermediate $\mu$ \citep{pierrehumbert23,benneke24}. On this reading the
gas-dwarf/water-world distinction is ill-posed for temperate sub-Neptunes
independently of formation location, and the observed $\mu$ reports the bulk
envelope composition. Whether that assumption holds is now contested for
exactly the planets in question. Coupling evolution models to \emph{ab
initio} hydrogen--water phase diagrams, \citet{howard25} find that demixing
may be ongoing in K2-18\,b, and that it would deplete water from the
observable layers and so account for the non-detection of H$_2$O in its
spectrum; they find partial depletion for TOI-270\,d.
\citet{piauletdemix25} reach a compatible conclusion from an
interior--atmosphere inference framework, identifying a demixing window for
bulk envelope metallicities of 100--700\,$Z_\odot$ --- the range inferred for
TOI-270\,d --- and affecting warm metal-rich sub-Neptunes between 330 and
500\,K generally. Both analyses are model-dependent and rest on phase
diagrams extrapolated beyond the range of direct measurement, and the
miscible interpretation remains viable; but a mixed envelope can no longer be
assumed for this population. We therefore treat miscibility as one pathway
among the several listed here rather than as the default state of a
metal-enriched envelope.

\emph{Preferential escape.} Hydrodynamic loss removes hydrogen faster than
heavier species, enriching the residual envelope. Our escape analysis places
this route within reach of all four hybrid worlds
(Sect.~\ref{sec:4.3}), most strongly for GJ\,3090\,b and GJ\,1214\,b. For
GJ\,3090\,b two independent arguments apply: its integrated loss is the
largest in the sample, and it lies below the boil-off threshold, where escape
is not bounded by Eq.~(\ref{eq:elim}) and a helium outflow is observed.
Either way the mechanism produces high $\mu$ from a primordially
hydrogen-rich planet.

It is worth recording that this pathway is anticipated within the gas-dwarf
framework itself. Discussing a planet in the radius valley,
\citet{rogers26} observe that such objects are expected to be in the final
stages of significant atmospheric loss and may retain thin, high mean
molecular weight atmospheres sculpted by mass loss and interior chemistry.
That is the hybrid mechanism, stated by the framework whose classification
criterion excludes the planets it produces. The disagreement is not about the
physics but about whether the resulting objects warrant a name.

\emph{Dry accretion of volatiles.} A planet formed interior to the ice line
may still acquire a carbon-rich, high-$\mu$ envelope. LP\,791-18\,c, with
CH$_4$/CO$_2 > 12$ and $Z \approx 316\,Z_\odot$, is the clearest example:
high $\mu$ produced without water.

These pathways are not mutually exclusive, and the resulting class is
continuous in $\mu$ rather than clustered. That is itself informative: a
bimodal composition distribution is a prediction of the dichotomy, and it is
not observed.

\subsection{The classification is unstable at both ends}
\label{sec:5.5}

The evidence supporting membership of the two end-member classes is weaker, and weaker in a
systematic way, than the evidence supporting the intermediate group. The asymmetry is one of
detection versus non-detection.

Planets we classify as hybrid rest on positive detections: CH$_4$ at 4.5$\sigma$ with an inferred
metallicity of $\approx$ 316 $Z_\odot$ for LP 791-18 c, CH$_4$ with CO$_2$ and CS$_2$ for TOI-270 d, a phase curve and
panchromatic spectrum indicating $Z \gtrsim 10^{3}\,Z_\odot$ for GJ 1214 b. Something is measured, and the measured
quantity places $\mu$ above 3.8 amu.

Planets classified at either extreme rest on weaker inference. At the high-$\mu$ end, an inferred
metallicity from a muted spectrum can be mimicked by high-altitude aerosols, and where an
escaping exosphere is observed the reservoir argument of \citet{rogers26} bounds $\mu$ from
above; this is what removes TOI-776 c from the sample's hybrid group (Sect.~\ref{sec:4.1}). At the low-$\mu$
end, the classification rests on the non-detection of water. For LTT 3780 c the detected CH$_4$
establishes an H$_2$-rich background but leaves the water abundance nominally constrained, so that a
volume mixing ratio above 8.4 per cent would carry the planet across the threshold. For K2-18 b the situation is more acute still, because the
published determinations straddle the threshold. The retrieval of \citet{madhu23},
which we adopt in Table~\ref{tab:atmos}, gives $\mu \approx$ 3.1 amu with water undetected. \citet{hu25}, from four
further NIRSpec transits, establish a water-rich composition requiring at least 10 per cent H$_2$O
by volume in the envelope, which corresponds to $\mu \ge$ 3.9 amu --- above the gas-dwarf boundary.
A reanalysis of the earlier data by \citet{schmidt25} reaches the opposite conclusion,
favouring an oxygen-poor mini-Neptune, while other work interprets the same spectrum as a
gas-rich envelope over a magma ocean. The classification of K2-18 b therefore depends on which
analysis is adopted, and the water abundance at which it changes class, 9.3 per cent by volume,
lies inside the range currently under dispute. Its lowest-in-sample equilibrium temperature
compounds this, since water condensed below the observable region would bias any retrieved $\mu$
downward.

Neither planet assigned to the gas-dwarf class therefore has a $\mu$ determination that stays on one
side of the threshold: LTT 3780 c spans it because water is unconstrained, K2-18 b because the
published analyses disagree. Both end-members are populated by classifications that current data
support only provisionally, while the intermediate group is populated by planets with positive compositional
detections. We do not take this as licence to reassign the provisional cases: an unconstrained $\mu$
is not evidence for a high one, and absorbing ambiguous planets into a preferred class is the
practice this paper criticises. The point is narrower and, we think, more consequential. A
taxonomy whose two categories are occupied mainly by objects that cannot yet be securely assigned
to them, while the planets with the firmest compositional measurements fall between the two, is
not partitioning the population usefully. That is an argument about the framework rather than
about any individual planet, and it does not require winning any of the individual disputes.

\subsection{Relation to existing terminology}
\label{sec:5.6}

The combination of escape observations with spectroscopic composition is not new to this work.
\citet{rogers26} constrain $\mu$ from above using observed mass-loss rates, applying the method
to four planets selected by the availability of a detected exosphere. Our approach is
complementary in three respects: our sample of eight is selected on physical criteria rather than
by the availability of a difficult observation, and is correspondingly more representative of the
M-dwarf population; we employ the restricted Jeans parameter as a retention filter, which their
analysis does not use; and our escape calculation integrates loss over the system lifetime to ask
whether a primordial envelope could have been removed, whereas theirs uses the instantaneous rate
to bound the present reservoir. The two constraints act in opposite directions on $\mu$ and are
jointly informative, as Sect.~\ref{sec:4.1} illustrates for GJ 3090 b.

Elements of this class have been named before. "Miscible envelope" \citep{benneke24} and
"mixed envelope" \citep{rogers25} describe individual planets or model grids in which hydrogen and
heavy volatiles coexist, and \citet{piaulet24} emphasise compositional diversity
among small sub-Neptunes. Our contribution is not the observation that intermediate compositions
exist, but three claims about them: that membership can be defined operationally by the two
criteria of Sect.~\ref{sec:5.1}; that the class is demographically dominant in the M-dwarf JWST sample
rather than a transition zone between well-populated end-members; and that the classification
survives being carried out with the gas-dwarf models' own threshold, so it cannot be attributed
to a choice of boundary favourable to either side of the debate.

\subsection{Beyond M dwarfs, and the difficulty of young systems}
\label{sec:5.7}

The two K-dwarf sub-Neptunes excluded by our stellar criterion both fall in
the hybrid regime. GJ~9827~d has $\mu = 9.8$--18~amu with H$_2$O as a
background gas at volume mixing ratio $> 31$ per cent and an inferred water
mass fraction of $32 \pm 10$ per cent
\citep{piaulet24} --- elevated, but well below the 50 per cent of
the 1:1 specification, and therefore a hybrid on criterion (ii) despite a
$\mu$ in the nominal water-world range. TOI-836~c has a muted spectrum
consistent with enrichment. Their inclusion would strengthen rather than
weaken the result, and suggests that the hybrid class is not a peculiarity
of M-dwarf systems. We exclude them here to keep the host population
homogeneous, and note the extension as a prediction.

A second boundary of the sample is stellar age, and it bears directly on how
our approach relates to the alternative. The gas-dwarf models discriminate
most sharply not through composition but through demography: because a
low-$\mu$ envelope contracts substantially as it cools while a volatile-rich
interior does not, the two classes are predicted to diverge in radius at ages
below $\sim$100\,Myr and to converge thereafter \citep{rogers25}. That route
requires young systems, and young systems are the regime in which
spectroscopic characterisation is compromised by the host itself.

Unocculted spots and faculae make the transit chord unrepresentative of the
disc-averaged stellar spectrum, imprinting a wavelength-dependent signal on
the transmission spectrum. The effect --- the transit light source effect,
formalised by \citet{rackham18,rackham19} and now the standard object of
mitigation in atmospheric retrievals --- scales with the covering fraction
and temperature contrast of the active regions, and therefore with stellar
activity. Its amplitude can exceed that of a genuine planetary atmospheric
feature in systems with active M and K-type hosts, so that the correction is
not a refinement applied to a detection but a determinant of whether there is
one \citep{rackhamdewit24}.

Two features of the effect make it particularly awkward for the question at
issue here. The first is that it is degenerate with the planetary quantities
we are trying to measure. \citet{iyerline20} show, for sub-Neptunes around
M dwarfs specifically, that unocculted heterogeneity stretches the
transmission spectrum in a manner degenerate with atmospheric scale height,
and that neglecting the correction biases the inferred planetary properties;
the bias is not removed by better precision, because the ability to
marginalise over the contamination is limited by the fidelity of the stellar
models themselves.

The second is that the false positives fall preferentially in water. The
photospheres of cool stars contain the same molecules sought in the planetary
atmosphere, H$_2$O among them, so that contamination can imprint a feature at
the wavelength of the species whose abundance our classification turns on
\citep{rackham18}. The consequence is visible in the published record: the
water features reported for GJ~486~b and GJ~1132~b remain of ambiguous origin
\citep{rackham23}, and for TRAPPIST-1~c two visits consistent with one
another at the 1.1$\sigma$ level were nonetheless found to be explicable
entirely by stellar contamination \citep{radica25}. Consistency between
visits, the usual argument for a planetary origin, was not sufficient.

The relevance is not to our census, whose hosts are old field M dwarfs, but
to the alternative route. Radius-based demography requires the age regime in
which this systematic is strongest, and composition-based classification of
mature systems does not carry it; conversely, demography of young systems
does not carry the escape-history ambiguity of Sect.~\ref{sec:4.3}, which is
what limits our own inference. The two routes are therefore complementary in
target regime rather than competing on the same objects, and each is weakest
where the other is strongest.

\subsection{Where a second water world would be found}
\label{sec:5.8}

The classification developed here implies a specific observational target profile, and it is not
the one that intuition suggests. A low $\Lambda$ does not indicate a water world: it indicates that the
planet cannot retain hydrogen, so that a high $\mu$ measured today is the expected residue of escape
and the origin is unrecoverable. Such planets fall in the hybrid class by criterion (ii) of
Sect.~\ref{sec:5.1}, whatever their $\mu$. The signature of a water world is the opposite combination --- high
$\Lambda$, so that escape is excluded as an explanation, together with a high measured $\mu$. It is the
combination that identifies LHS 1140 b ($\Lambda$ = 109, $\mu \approx$ 28 amu), and it requires a cold, massive,
strongly bound planet.

Among M-dwarf sub-Neptunes with measured masses that lack JWST spectra, we find no object with
that profile. The available candidates --- TOI-1201 b ($\Lambda$ = 28), K2-3 b ($\Lambda$ = 43), TOI-269 b
($\Lambda$ = 45) --- all lie in the intermediate retention regime and at equilibrium temperatures of
500--700 K. For these our framework predicts elevated $\mu$, but with an origin that escape will not
resolve: they are predicted hybrids, not water-world candidates. The prediction is falsifiable
in the useful direction, since a low-$\mu$ H$_2$ envelope in any of them would contradict it.

The absence of a second LHS 1140 b analogue in the accessible sample is itself informative, and
it is at least partly a selection effect rather than a statement about the population. Kepler
identified few M-dwarf sub-Neptunes, and those it did are too faint for transmission
spectroscopy; interpreting a spectrum of a small planet moreover requires a mass measured to
better than 5$\sigma$ to break the degeneracy between surface gravity and mean molecular weight
(\citealt{batalha19}), a precision rarely reached by transit-timing masses in the Kepler sample.
The characterisable population is therefore drawn almost entirely from TESS, and within it from
the nearest and brightest systems. Extending the search for strictly volatile-rich worlds means
targeting cold, massive planets around nearby M dwarfs --- the regime where escape is weakest and
where a high $\mu$, if measured, cannot be explained away.

\subsection{Falsification}
\label{sec:5.9}

The class is falsifiable in three ways. Retrievals resolving $\mu$ for GJ 1214 b and GJ 3090 b above
the aerosol continuum would test whether their envelopes are enriched hydrogen or genuinely
water-dominated; values in the 6--18 amu range with corresponding water mass fractions would move
them to the water-world class. Interior modelling constraining water mass fractions for
TOI-270 d and LP 791-18 c would test condition (ii) directly. And a larger sample showing
bimodality in $\mu$, with a deficit at 5--10 amu, would indicate that the intermediate objects are a
transition zone rather than a population. Present data show no such deficit.

For GJ~3090~b specifically the falsifying measurement is not a better metallicity bound but a constraint on the aerosol pressure, since that is the quantity selecting between the two branches of \citet{parker25}.

\subsection{Limitations}
\label{sec:5.10}

Four limitations bound the result.

\emph{Sample size and selection.} Eight planets is a small sample, and it is not drawn from a
volume-limited or otherwise controlled parent population. JWST targets are chosen for
transmission-spectroscopy favourability, which preferentially selects low-density planets with
bright hosts, and the sample is therefore biased towards extended envelopes. The direction of
this bias works against our result --- a selection favouring puffy, low-$\mu$ atmospheres should
over-represent gas dwarfs, and we find the opposite --- but it prevents the fractions we quote from
being read as occurrence rates. The claim is that hybrid worlds dominate the characterised
sample, not that we have measured their frequency in the underlying population.

\emph{Compositional gradients.} The inference from atmospheric $\mu$ to bulk composition assumes the
envelope is well mixed. If hydrogen and water demix, compositional gradients develop and the link
between bulk envelope metallicity and that of the observable atmosphere is broken
\citep{piauletdemix25}, so that a measured $\mu$ constrains only the layers probed. Published
models place two planets in our sample in this regime: K2-18 b, where demixing would deplete
water from the observable atmosphere, and TOI-270 d, where the inferred envelope metallicity
falls inside the demixing window (\citealt{howard25}; \citealt{piauletdemix25}). This is the
most serious challenge to the inference chain of Sect.~\ref{sec:3.1}, and we do not consider it resolved.

We note, however, that it bears on the gas-dwarf threshold with equal force. A criterion stated
in atmospheric $\mu$ but used to assign planets to classes defined by bulk composition --- rocky core
with H/He, or 1:1 rock-ice interior --- presupposes exactly the mixing that these models question.
If the presupposition fails, the failure is in the classification scheme rather than in any
particular application of it, which is the argument of Sect.~\ref{sec:5.1} arrived at from the interior
side.

\emph{Atmospheric versus bulk composition.} Transmission spectra probe millibar pressures. The
inference from an observed $\mu$ to a bulk volatile inventory requires that the sampled layer be
representative, and compositional gradients, condensation, or photochemical processing can break
that assumption. This is most acute for the coldest object: at Teq $\approx$ 255 K a fraction of K2-18 b's
water may be condensed below the photosphere, which would bias its retrieved $\mu$ downward and its
classification towards the gas-dwarf end. Our criterion (ii) mitigates this by requiring interior
evidence, but that evidence is itself model-dependent.

\emph{The aerosol degeneracy.} For GJ 1214 b, GJ 3090 b and TOI-776 c, $\mu$ is bounded rather than
retrieved. High-altitude aerosols and high mean molecular weight both mute spectral features, and
although the enrichment of the first two is independently supported --- by phase-curve amplitude
for GJ 1214 b, by a subdued helium outflow for GJ 3090 b --- their placement within the hybrid
range rests on metallicity bounds, not on resolved $\mu$.

\emph{Proximity to the threshold.} Because $\mu$ is a model-dependent inference rather than a measurement
(Sect.~\ref{sec:3.1}), planets whose retrieved value lies close to 3.8 amu could change class under a
different retrieval. Two are in this position. TOI-270 d, at $\mu \approx$ 5.0--5.8 amu, sits 1.2--2.0 amu
above the threshold; K2-18 b, at $\mu \approx$ 3.1 amu, sits 0.7 amu below it. For these the classification
should be regarded as provisional on the retrieval framework adopted in the source analyses. The
remaining six are separated from the threshold by margins that no plausible systematic in the
retrieval would close: LHS 1140 b by roughly 24 amu, LTT 3780 c by 1.3 amu with an independently
detected hydrogen-dominated composition, and the rest by metallicity bounds rather than by
retrieved central values.

\emph{Escape assumptions.} Equation (2) carries the standard uncertainties of the energy-limited
approach: $\eta$ varies with gravitational potential and irradiation, the past XUV history is inferred
from activity--age relations rather than measured, and several stellar ages are poorly constrained.
Detailed hydrodynamic models capture physics the formulation omits. We nonetheless prefer a
homogeneous analytic treatment for a comparative sample, because applying heterogeneous published
models --- computed with different codes, efficiencies and XUV prescriptions --- would confound
planetary properties with differences in code architecture, and it is the differential
susceptibility of the planets that our classification requires. The absolute value of X is
assumption-dependent; the rank ordering of the sample, and in particular the segregation of
LHS 1140 b by two to three orders of magnitude, is not. We have restricted origin claims to cases
where the margin exceeds the assumptions, and the classification of the intermediate objects as
indeterminate reflects the limits of this layer rather than a positive finding.

\section{Conclusions}
\label{sec:6}

We have classified the eight M-dwarf sub-Neptunes with constrained
atmospheric mean molecular weights against the gas-dwarf threshold adopted
within the gas-dwarf literature, and asked of those above it whether their
present composition requires a volatile-rich origin.

Four of the eight belong to neither class. Their envelopes, spanning
$3.8 < \mu \le 18$\,amu, are accounted for by magma-ocean equilibration,
supercritical miscibility, preferential hydrogen escape, or accretion
interior to the water-ice line --- processes that produce an elevated mean
molecular weight without requiring formation as a volatile-dominated body. We
propose \emph{hybrid worlds} as the designation for this class, defined by a
mean molecular weight above the gas-dwarf threshold together with a present
composition that does not require a volatile-rich origin. They are the
commonest outcome in this sample, outnumbering both end-members together.

The two end-members are realised, but sparsely. One planet, LHS\,1140\,b,
requires a volatile-rich origin: it could not have shed a primordial envelope
over its lifetime, and it has none, which places its volatile inventory at
formation. Its water mass fraction of 9--19 per cent \citep{cadieux24} is
well below the 1:1 idealisation and does not bear on the classification,
which follows from origin rather than from present bulk composition. At the
other end, one planet, TOI-776\,c, returns tentatively to the gas-dwarf side
once the constraint from its observed escape is applied, though we stress
that the combined limit of \citet{rogers26}, $\mu \le 12.4$\,g\,mol$^{-1}$,
lies inside the hybrid range: what their analysis breaks is the aerosol
degeneracy, and they call the result tentative.

The two objects we have not counted --- K2-18\,b and LTT\,3780\,c --- do not
fall below the boundary but span it, the first because the published
determinations disagree across it, the second because the water abundance is
unconstrained. Neither can be assigned to the gas-dwarf class on present
data, and either could prove hybrid. The minimum hybrid count in this sample
is therefore four and the maximum gas-dwarf count is three, on a threshold
supplied by the gas-dwarf literature itself.

Three features of the result matter more than the count.

The first is that the gas-dwarf class, at its own definition, is empty of
firm members. No planet in the sample has a $\mu$ retrieved from detected
molecular features that places it below 3.8\,amu, and no planet displays a
solar-composition hydrogen envelope; the two lowest-$\mu$ objects are
themselves enriched relative to solar. The single planet we assign to the
class is assigned by an escape-reservoir argument applied to a featureless
spectrum, not by a measurement.

The second is that the selection function runs against the result. JWST
targets are chosen for transmission-spectroscopy favourability, which
preferentially selects low-density planets with extended atmospheres around
bright hosts. Such a selection should over-represent puffy, low-$\mu$
envelopes --- that is, gas dwarfs. We find the opposite. The bias prevents
the fractions we quote from being read as occurrence rates, and we do not
read them that way; but a bias pushing in the opposite direction to the
result is evidence for it, not against.

The third is the asymmetry of the evidence, which points the same way. The
hybrid assignments rest on positive detections: CH$_4$ at 4.5$\sigma$ with an
inferred metallicity of $\approx 316\,Z_\odot$ for LP\,791-18\,c, and CH$_4$
with CO$_2$ and CS$_2$ for TOI-270\,d. Something is measured, and the
measured quantity places $\mu$ above the boundary. The assignments at both
end-members rest on non-detections, on disputed retrievals, or on featureless
spectra. A taxonomy whose two categories are occupied mainly by objects that
cannot yet be securely assigned to them, while the planets with the firmest
compositional measurements fall between the two, is not partitioning the
population usefully.

We have tested whether the census survives the range of thresholds the
gas-dwarf framework itself proposes, and the answer is instructive rather
than simply reassuring. Read as a count it does not: four planets satisfy
neither end-member at 3.8\,amu \citep{calder26}, three at the $\approx 5$\,amu
reached by magma-ocean equilibration \citep{rogers25}, and one at the 10\,amu
upper bound that outgassing thermodynamics permits \citep{heng25}. But the
collapse at the high threshold is not a vindication of the dichotomy. At a
10\,amu boundary the gas-dwarf class absorbs LP\,791-18\,c, whose molecular
ratios place its accretion interior to the ice line, and TOI-270\,d,
described by its discovery analysis as a miscible envelope rather than a
layered one. Neither is a rocky core beneath retained primordial H/He, which
is what the term denotes. There is no threshold in $\mu$ that leaves both
end-member classes with their physical content intact while emptying the
region between them: where the boundary is low that region is populated, and
where it is high the gas-dwarf class becomes a residual bin containing
objects its own formation model does not produce.

The escape analysis on which the origin classification rests has been
recomputed for this work with an XUV history taken from M-dwarf
activity--age relations at each host's spectral type \citep{engle24} rather
than parameterised as a free enhancement factor, and with the conversion from
measured X-ray luminosities to the 1--912\,\AA\ band made explicit
\citep{sanzforcada11}. The partition it produces at $X = 1$ is what the
classification uses, and it is robust under the propagated uncertainties; the
fine ordering within the susceptible group is not, and we read no order into
it. The classification is also consistent with the independent upper limit
that an observed escaping exosphere places on $\mu$ \citep{rogers26},
including for GJ\,3090\,b, the only planet in our sample with both a detected
outflow and a high inferred $\mu$.

Neither end-member model is refuted. Both are realised, and both are minority
outcomes. What the sample does not support is the framing: a debate between
two classes, conducted over a population in which the commonest object
belongs to neither.

Three measurements would sharpen this. Resolved retrievals for the
aerosol-obscured planets would establish whether their enrichment is
hydrogen-based or volatile-based; for GJ\,3090\,b specifically what is needed
is a constraint on the aerosol pressure, the quantity that selects between
the two branches of \citet{parker25}. Interior modelling constraining water
mass fractions for LP\,791-18\,c and TOI-270\,d would test our second
criterion where it currently rests on molecular ratios alone. Direct
constraints on the past XUV histories of the host stars would reduce the
dominant remaining uncertainty in the escape analysis. All three are within
reach of current facilities.

The extension that would settle the matter is demographic. Our sample is set
by which planets have been observed, not by any statistical design. A
homogeneously selected sample spanning the M-dwarf sub-Neptune regime, large
enough to test whether $\mu$ is distributed continuously or bimodally between
3.8 and 18\,amu, would distinguish a genuine intermediate population from a
transition zone, and would convert the statement made here about a
characterised sample into one about occurrence. Present data show no deficit
at intermediate values; a larger sample that did would refute the class
proposed here.

\begin{table*}
\centering
\caption{System and planet properties of the sample, ordered by decreasing atmospheric mean
molecular weight. $T_{\rm eff}$ is the host effective temperature and $T_{\rm eq}$ the planetary
equilibrium temperature.}
\label{tab:sample}
\begin{tabular}{l c c c c c l}
\toprule
Planet & SpT & $T_{\rm eff}$ (K) & $T_{\rm eq}$ (K) & $R_p$ ($R_\oplus$) & $M_p$ ($M_\oplus$) & Reference \\
\hline
LHS 1140 b  & M4.5V & 3096 & 226 & $1.730\pm0.025$ & $5.60\pm0.19$ & \citet{cadieux24} \\
GJ 1214 b   & M4.5V & 3250 & 596 & $2.733\pm0.033$ & $8.41\pm0.36$ & \citet{cloutier19} \\
LP 791-18 c & M6V   & 2960 & 355 & $2.44\pm0.10$   & $7.10\pm0.70$ & \citet{peterson23} \\
GJ 3090 b   & M2V   & 3556 & 693 & $2.18\pm0.06$   & $4.52\pm0.47$ & \citet{lamontagne26} \\
TOI-776 c   & M1V   & 3275 & 415 & $2.047\pm0.081$ & $6.90^{+2.6}_{-2.5}$ & \citet{fridlund24} \\
TOI-270 d   & M3V   & 3506 & 387 & $2.13\pm0.06$   & $4.78\pm0.43$ & \citet{vaneylen21} \\
K2-18 b     & M2.5V & 3457 & 255 & $2.61\pm0.09$   & $8.63\pm1.35$ & \citet{cloutier19} \\
LTT 3780 c  & M3.5V & 3331 & 353 & $2.39\pm0.08$   & $8.04\pm0.90$ & \citet{bonfanti24} \\
\hline
\end{tabular}
\end{table*}

\begin{table*}
\centering
\caption{Atmospheric constraints and derived quantities. The quality flag on $\mu$ is R (retrieved
from detected features), I (inferred from a bounded metallicity in a muted spectrum), T (tentative,
flat spectrum), C (conditional: the limit holds only under an assumed aerosol pressure),
H (hydrogen background established, enrichment unconstrained) or D (published
determinations disagree across the threshold). $\Lambda$ is the
restricted Jeans parameter recomputed from Eq.~(\ref{eq:lambda}); $X$ is the ratio of removable
H/He mass to primordial envelope mass, quoted as the median and 16--84 per cent interval of the
Monte Carlo of Sect.~\ref{sec:4.3} and including the per-planet tidal factor
$K$. Classification follows Sect.~\ref{sec:5.1}.}
\label{tab:atmos}
\begin{tabular}{l l c r l c l l}
\toprule
Planet & $\mu$ (amu) & Flag & $\Lambda$ & $X$ & $P(X{>}1)$ & JWST detection & Class \\
\hline
LHS 1140 b  & $\approx28$              & R & 108.6 & 0.07 (0.02--0.18) & 0.5\% & H$_2$-rich disfavoured; N$_2$-dom. & Water world \\
GJ 1214 b   & $>15$                    & I &  39.1 & 7.0 (2.5--20)     & 97\%  & muted; $Z\sim10^{3}\,Z_\odot$      & Hybrid \\
LP 791-18 c & $7.68^{+2.24}_{-1.48}$   & R &  62.1 & 2.2 (0.7--7.4)    & 75\%  & CH$_4$; hazes; $Z\approx316\,Z_\odot$ & Hybrid \\
GJ 3090 b   & $>7.1^{\rm e}$           & C &  22.7 & 16 (4.6--54)$^{\rm d}$ & 99\% & metal-enriched; He outflow    & Hybrid$^{\rm f}$ \\
TOI-776 c   & $\approx6$--8            & T &  60.9 & 0.55 (0.16--1.83) & 31\%  & flat                                & Gas dwarf$^{\rm a}$ \\
TOI-270 d   & 5.0--5.8                 & R &  43.9 & 3.8 (1.1--13)     & 86\%  & CH$_4$ + CO$_2$ + CS$_2$            & Hybrid \\
K2-18 b     & 3.1 (disputed)           & D &  98.3 & 0.06 (0.02--0.14) & 0.0\% & CH$_4$ + CO$_2$; H$_2$-rich         & Ambiguous$^{\rm b}$ \\
LTT 3780 c  & 2.5--5.6                 & H &  72.2 & 0.41 (0.14--1.16) & 20\%  & CH$_4$ ($\sim$1\%); H$_2$-rich      & Ambiguous$^{\rm c}$ \\
\hline
\multicolumn{8}{l}{$^{\rm a}$ Returned tentatively to the gas-dwarf class following the joint escape-plus-spectroscopy analysis of \citet{rogers26}; their}\\
\multicolumn{8}{l}{\phantom{$^{\rm a}$ }combined limit, $\mu\le12.4$ g\,mol$^{-1}$, lies inside the hybrid range and does not exclude membership. See Sect.~\ref{sec:4.4}.}\\
\multicolumn{8}{l}{$^{\rm b}$ \citet{madhu23} retrieve $\mu\approx3.1$, while \citet{hu25} require $\ge$10 per cent H$_2$O by volume, implying $\mu\ge3.9$; the permitted}\\
\multicolumn{8}{l}{\phantom{$^{\rm b}$ }range spans the threshold (Sect.~\ref{sec:4.1}). Its retained envelope is secure, its $\mu$ is not.}\\
\multicolumn{8}{l}{$^{\rm c}$ CH$_4$ establishes an H$_2$ background, but H$_2$O is only nominally constrained and the permitted range spans the}\\
\multicolumn{8}{l}{\phantom{$^{\rm c}$ }threshold; envelope retained at $\approx$80 per cent confidence.}\\
\multicolumn{8}{l}{$^{\rm d}$ Upper bound: at this instellation the outflow is expected to be radiation--recombination limited and Eq.~(\ref{eq:elim})}\\
\multicolumn{8}{l}{\phantom{$^{\rm d}$ }overestimates $\dot M$ (Sect.~\ref{sec:4.3}).}\\
\multicolumn{8}{l}{$^{\rm e}$ From CRIRES+ high-resolution spectroscopy \citep{parker25}, not from the JWST spectrum, and conditional on an}\\
\multicolumn{8}{l}{\phantom{$^{\rm e}$ }aerosol deck $\lesssim10^{-2}$ bar; the alternative branch leaves the metallicity unconstrained (Sect.~\ref{sec:2.3}).}\\
\multicolumn{8}{l}{$^{\rm f}$ Provisional. Criterion (ii) is satisfied robustly, by the escape history and by $\Lambda$ below the boil-off threshold;}\\
\multicolumn{8}{l}{\phantom{$^{\rm f}$ }it is criterion (i) that is conditional here. \citet{parker25} state that a water world cannot be excluded.}\\
\end{tabular}
\end{table*}

\begin{figure}
\centering
\includegraphics[width=\columnwidth]{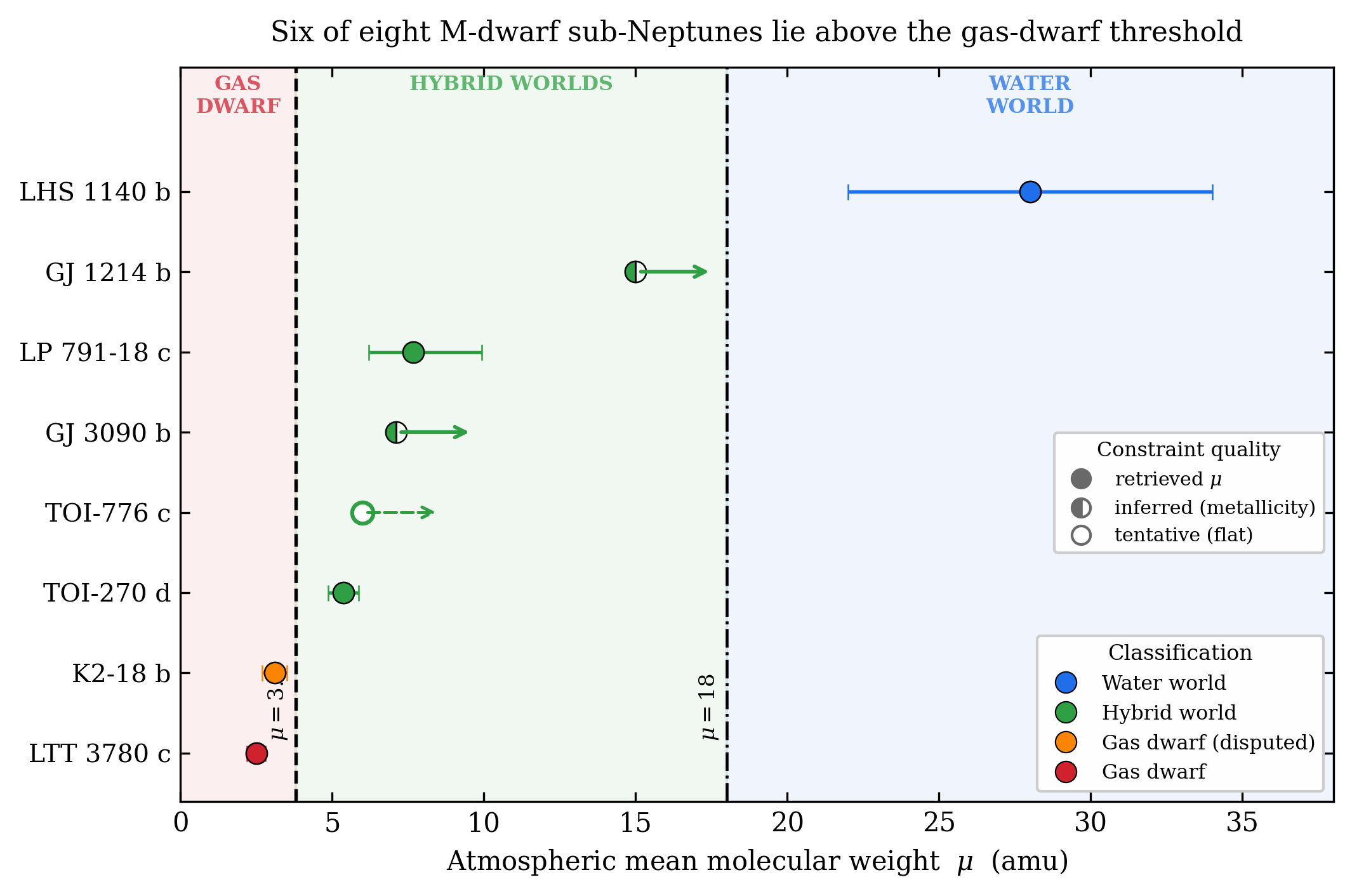}
\caption{Atmospheric mean molecular weight of the eight M-dwarf sub-Neptunes of Table~\ref{tab:sample}, ordered by $\mu$. The vertical dashed line marks $\mu$ = 3.8 amu, the upper limit adopted for the gas-dwarf class by the gas-dwarf models themselves \citep{calder26}; the dash-dotted line marks $\mu$ = 18 amu, the value for pure H$_2$O, above which hydrogen cannot be a significant envelope constituent. Shading indicates the three composition classes these bounds define: gas dwarf (red), hybrid world (green) and water world (blue), with point colours following the same scheme. Marker fill encodes the quality of the $\mu$ constraint: filled symbols with error bars are retrieved from detected molecular features; half-filled symbols with arrows are lower limits inferred from a bounded atmospheric metallicity in a muted spectrum; the open symbol (TOI-776 c) is a tentative lower limit from a flat spectrum in which the aerosol/high-$\mu$ degeneracy is unresolved. Six of the eight planets have a permitted range lying wholly above the gas-dwarf threshold and none lies wholly below it; the two remaining objects span the boundary (Sect.~\ref{sec:4.1}).}
\label{fig:1}
\end{figure}

\begin{figure}
\centering
\includegraphics[width=\columnwidth]{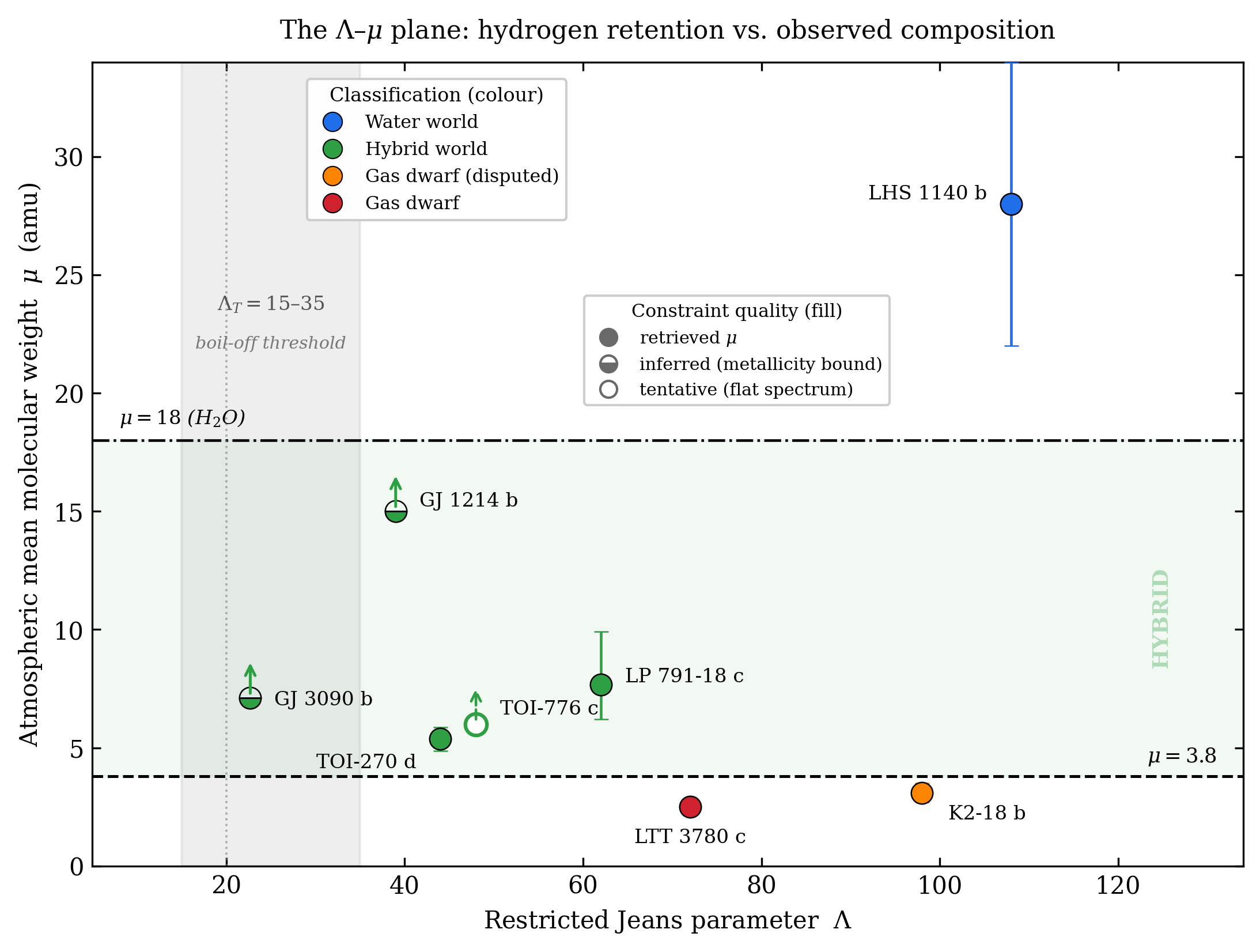}
\caption{The $\Lambda$--$\mu$ plane. The restricted Jeans escape parameter $\Lambda$ \citep{fossati17}, recomputed homogeneously for all eight planets from consistently sourced Mp, Rp and Teq, is plotted against the measured or bounded atmospheric mean molecular weight. Symbols, colours and fills are as in Fig.~\ref{fig:1}. The grey band spans $\Lambda_T$ = 15--35, the range within which \citet{fossati17} locate the boil-off threshold depending on system parameters; the dotted line marks $\Lambda_T$ = 20, the value corresponding to the $R_p$/$R_{\rm B}$ = 0.1 criterion of \citet{owenwu16} in the isothermal limit. Horizontal lines and green shading reproduce the class boundaries of Fig.~\ref{fig:1}. Only GJ~3090~b lies below the threshold band, and it does so for both the discovery mass of \citet{almenara22} and the revised mass of \citet{lamontagne26}: at its $T_{\rm eq}$ and mass the M2 grid of \citet{fossati17} implies a threshold in the middle of the $\Lambda_T\approx21$--30 range appropriate to M-dwarf hosts, above both values. Helium escape is directly detected for this planet. The remaining seven lie above the band under any choice of $\Lambda_T$ within the quoted range, so the partition is insensitive to where the threshold is placed. Among those seven, LHS 1140 b ($\Lambda$ = 109), K2-18 b ($\Lambda$ = 98) and LTT 3780 c ($\Lambda$ = 72) share a retention regime but differ in $\mu$ by an order of magnitude: where hydrogen could have been retained, only the spectrum determines whether it was.}
\label{fig:2}
\end{figure}

\begin{figure}
\centering
\includegraphics[width=\columnwidth]{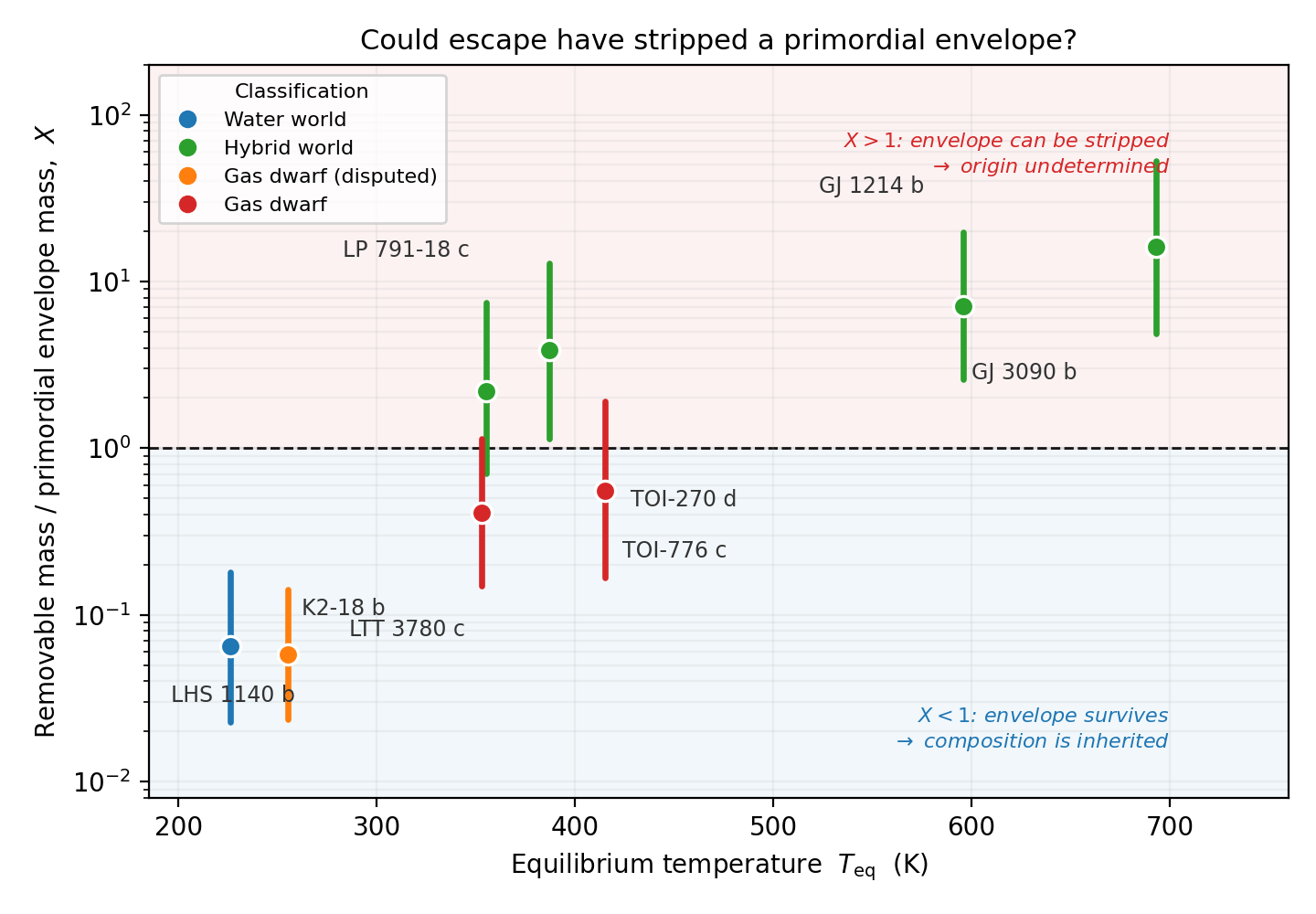}
\caption{Ratio $X$ of the H/He mass removable by escape over the system lifetime to the mass of a
plausible primordial envelope, against equilibrium temperature. Markers give the median of the
Monte Carlo of Sect.~\ref{sec:4.3} and bars the 16--84 per cent interval, propagating the heating
efficiency, the initial envelope mass fraction, the scatter of the X-ray to EUV conversion and,
where $L_{\rm bol}$ is inferred rather than published, the Bond albedo. The integrated XUV history
is taken from the M-dwarf activity--age relations of \citet{engle24} at each host's spectral type
rather than parameterised. The dashed line marks $X=1$: above it (red) escape could have removed
the envelope entirely, so the present composition need not be primordial; below it (blue) the
envelope survives and the observed composition is inherited. K2-18~b and LHS~1140~b lie below the
line decisively, with $P(X>1)<1$ per cent, and LTT~3780~c follows at 20 per cent. GJ~3090~b is the
most susceptible planet in the sample; its energy-limited rate exceeds that inferred from its
helium absorption by two orders of magnitude, as expected where the outflow becomes
radiation--recombination limited, and its $X$ should be read as an upper bound. Colours as in
Fig.~\ref{fig:1}.}
\label{fig:3}
\end{figure}

\begin{figure}
\centering
\includegraphics[width=\columnwidth]{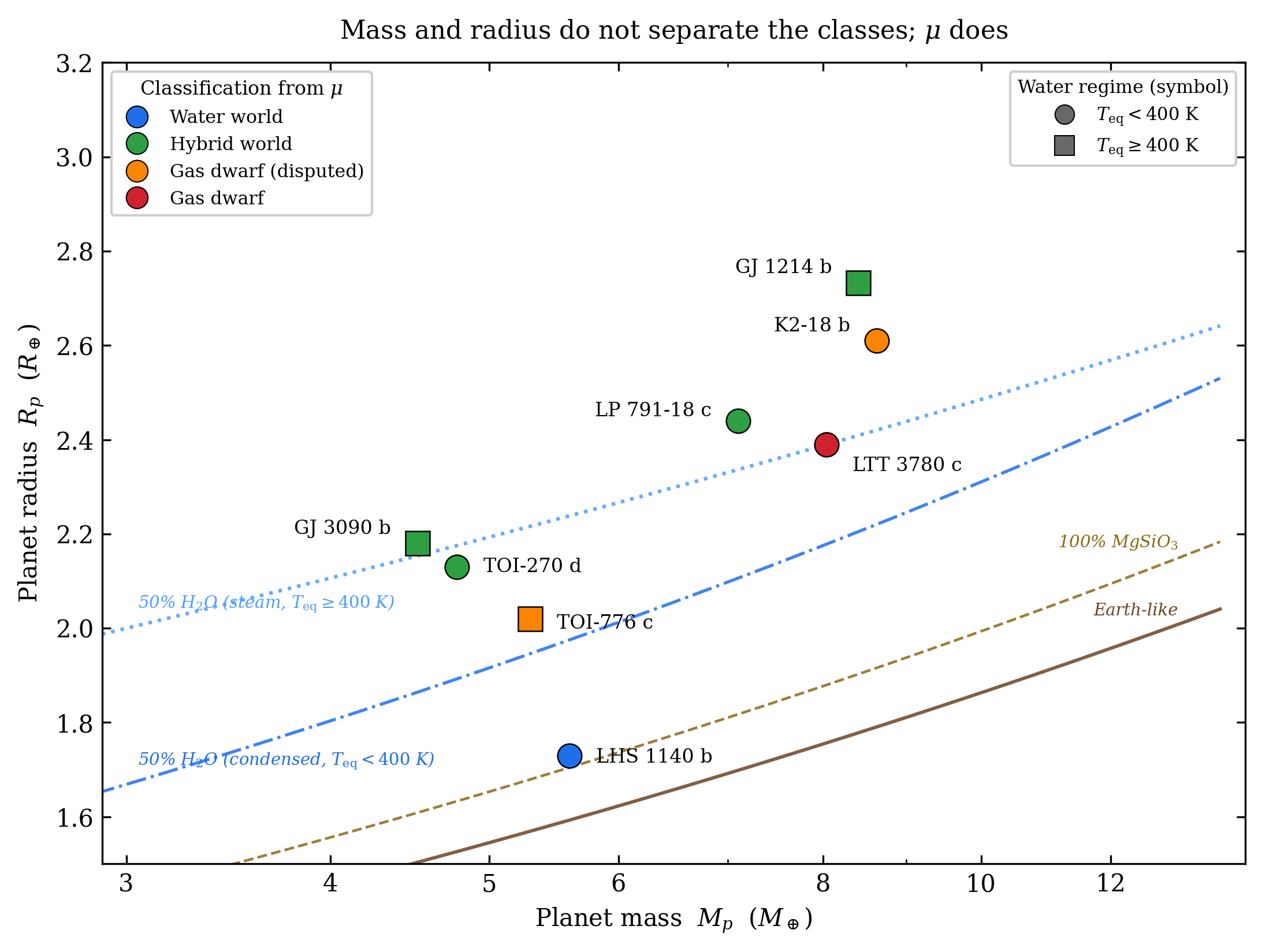}
\caption{Mass--radius diagram of the sample, with point colours giving the composition class assigned from $\mu$ (Fig.~\ref{fig:1}) and symbol shape distinguishing the two water regimes: circles for $T_{\rm eq}$ < 400 K, where a water layer is condensed, squares for $T_{\rm eq}$ $\ge$ 400 K, where it is in a steam state and inflates the radius. Curves are the tabulated relations of \citet{zeng19} for an Earth-like composition (R/R$_\oplus$ = $M^{0.27}$) and for a 1:1 silicate-to-ice water world in the condensed regime (R/R$_\oplus$ = 1.24 $M^{0.27}$), the latter being the relation adopted by \citet{luque22} and quoted by \citet{rogers23}; the pure MgSiO$_3$ curve is from \citet{zeng16} and the steam 50 per cent H$_2$O curve (R/R$_\oplus$ = 1.64 $M^{0.18}$) from the fits of \citet{parc24}, which reproduce the Zeng et al. relations in the condensed regime. The figure makes the degeneracy explicit: LTT 3780 c and LP 791-18 c differ by less than 1 M$_\oplus$ and 0.05 R$_\oplus$ yet have $\mu$ of approximately 2.5 and 7.7 amu, while GJ 1214 b and K2-18 b are similarly close in mass and radius with opposite classifications. Mass and radius do not separate the classes; the spectroscopic mean molecular weight does. Note also that LHS 1140 b lies well below the condensed 50 per cent H$_2$O curve, its position implying a water mass fraction near 18 per cent, consistent with the 9--19 per cent derived by \citet{cadieux24} and far short of the 1:1 idealisation; as set out in Sect.~\ref{sec:5.1}, this does not affect its classification, which follows from origin rather than from present bulk composition.}
\label{fig:4}
\end{figure}

\section*{Acknowledgements}

This work is based on observations made with the NASA/ESA/CSA James Webb
Space Telescope, obtained from the Mikulski Archive for Space Telescopes at
the Space Telescope Science Institute, which is operated by the Association
of Universities for Research in Astronomy, Inc., under NASA contract
NAS 5-03127. This research has made use of the NASA Exoplanet Archive,
operated by Caltech under contract with NASA under the Exoplanet Exploration
Program, and of NASA's Astrophysics Data System. The analysis made use of
\texttt{numpy} \citep{harris20}, \texttt{scipy} \citep{virtanen20},
\texttt{astropy} \citep{astropy22} and \texttt{matplotlib} \citep{hunter07}.

\section*{Funding}

This research received no specific grant from any funding agency in the
public, commercial, or not-for-profit sectors.

\section*{Declaration of competing interest}

The author declares no known competing financial interests or personal
relationships that could have appeared to influence the work reported in
this paper.

\section*{Data availability}

The JWST transmission spectra used here are publicly available from the
Mikulski Archive for Space Telescopes. The derived quantities of Tables~1
and~2, and the code used to compute the restricted Jeans parameter, the
metallicity-to-mean-molecular-weight conversion, the energy-limited escape
estimates and the Monte Carlo sensitivity analysis, are available from the
author on reasonable request.

\section*{Declaration of generative AI and AI-assisted technologies in the
writing process}

During the preparation of this work the author used Claude (Anthropic) for
language editing and for translation between Spanish and English. These
tools were not used to generate scientific results, to perform or interpret
the analysis, or to select or evaluate the literature. After using them,
the author reviewed and edited the content as needed and takes full
responsibility for the content of the publication.

\appendix

\renewcommand{\thetable}{A\arabic{table}}
\setcounter{table}{0}
\renewcommand{\thefigure}{A\arabic{figure}}
\setcounter{figure}{0}

\section{Parameters and inputs of the escape calculation}
\label{app:escape}

Table~\ref{tab:escape} collects every quantity entering equations (1) and (2), so that the calculations can be
reproduced. Orbital separations and stellar masses are from the discovery or refinement papers
cited in Table~\ref{tab:sample}. Stellar ages are the values quoted in those sources; where an age is not well
constrained we adopt 5 Gyr, which is representative of field M dwarfs, and note that the escape
result depends on this choice linearly through the integration time. Three systems have ages that
are constrained and materially different from that default: GJ~3090 (1.02 Gyr, gyrochronology;
\citealt{almenara22}), K2-18 (2.9 Gyr), and TOI-776 (1--4 Gyr; \citealt{loyd25}), whose host has a
measured rotation period of 21.13 d. For four planets the present-day flux is anchored on a measured stellar X-ray luminosity, listed
in the final column of Table~\ref{tab:escape}. Where a measurement is an upper limit (K2-18, LTT 3780) we
adopt the limit, which makes the resulting removable mass fraction an upper bound and therefore
conservative with respect to our conclusion that these planets retain their envelopes. For the
remaining planets we scale $L_{\rm XUV}$ from the bolometric luminosity using an activity-calibrated
ratio appropriate to the spectral type and age, and we regard the resulting rates as order-of-
magnitude estimates only.

The integrated history is the dominant uncertainty and is not measured for any target. We adopt
the standard picture in which M dwarfs remain in the saturated regime, $L_{\rm XUV}$/$L_{\rm bol} \approx$ 10$^{-3}$, for
the first 1--2 Gyr and decay thereafter, and we parameterise the resulting integrated exposure as
an enhancement factor of 10--50 relative to the present-day flux. This range, rather than a single
value, is what generates the intervals in X quoted in Sect.~\ref{sec:4.1} and shown in Fig.~\ref{fig:3}. The Monte
Carlo of Sect.~\ref{sec:4.1} samples the enhancement from a log-normal distribution whose median is
varied between 15 and 60 to test sensitivity, together with $\eta$ from a log-normal of median 0.15
and 68 per cent interval 0.09--0.26, and the initial envelope mass fraction from a log-normal of
median 2 per cent truncated to 0.3--15 per cent.

Equation (2) is an analytic upper bound on thermally driven escape, not a hydrodynamic
simulation. It omits the detailed radiative and chemical structure of the upper atmosphere, the
transition between energy-limited and recombination-limited regimes, and any non-thermal loss.
Its value here is comparative: applied uniformly across the sample with consistently sourced
inputs, it ranks the planets by susceptibility to envelope loss, and it is that ranking, not the
absolute rates, on which the origin classification of Sect.~\ref{sec:4.1} depends. Where detailed
evolutionary models exist we use them as external checks rather than as inputs, and both agree
with our estimate in sign and margin: \citet{cadieux24} find that LHS 1140 b retains a
0.1 per cent envelope over 10 Gyr, against our removable fraction of $\approx$ 0, and published
modelling of K2-18 b finds its primordial envelope survives, against our 0.08--2.03 in X.

Three planets in the sample have a directly detected escaping exosphere: GJ 3090 b in
helium \citep{ahrer25}, TOI-776 c in Lyman-$\alpha$ \citep{loyd25}, and GJ 1214 b, for which
escape has been reported but the inferred rate is model dependent. For these, the requirement
that the observed loss be sustainable over the system lifetime places an upper limit on $\mu$
\citep{rogers26}, complementing the lower limits from spectroscopy. We reproduce their
worked example as a check on our implementation, recovering X $\ge$ 0.53 and $\mu \le$ 3.1 amu for a
5 $M_\oplus$ core with a 1 per cent envelope at 5 Gyr, and apply the constraint to GJ 3090 b in
Sect.~\ref{sec:4.1}.

\subsection*{A.1. From measured X-rays to the integrated XUV band}

Equation~(\ref{eq:elim}) requires the flux integrated over 1--912\,\AA. What is measured for the
anchored hosts is a soft X-ray luminosity in a narrower band, and the two differ by more than an
order of magnitude for stars of this activity level. We adopt the relation of
\citet{sanzforcada11},
\begin{equation}
\log L_{\rm EUV} = (4.80 \pm 1.99) + (0.860 \pm 0.073)\,\log L_{\rm X},
\label{eq:sf11}
\end{equation}
with $L_{\rm X}$ in 5--100\,\AA\ and $L_{\rm EUV}$ in 100--920\,\AA, and take
$L_{\rm XUV}=L_{\rm X}+L_{\rm EUV}$. Because the slope is below unity the EUV term dominates
increasingly towards low activity: for the hosts in this sample $L_{\rm EUV}/L_{\rm X}$ lies
between 11 and 15, so that adopting $L_{\rm X}$ itself as $F_{\rm XUV}$ would understate the escape
rate by an order of magnitude.

Two qualifications attach to this step. Equation~(\ref{eq:sf11}) was calibrated over
$L_{\rm X}\approx10^{26}$--$10^{31}$\,erg\,s$^{-1}$; GJ~1214 ($\log L_{\rm X}=25.87$) lies below
that range and LHS~1140 ($26.13$) sits at its lower edge, so for these two the conversion is an
extrapolation into the regime where the EUV excess is largest and least constrained. And the
relation was derived for dwarfs from F through M, with no M-dwarf-specific recalibration of
comparable coverage available; the values should be read as order-of-magnitude estimates,
consistent with the comparative use to which Sect.~\ref{sec:4.3} puts them.

One external check is available. \citet{lalitha14}, whose X-ray measurement anchors GJ~1214 in
Table~\ref{tab:escape}, apply the same conversion to that star and obtain
$\log L_{\rm XUV}=27.09\pm0.02$; our implementation returns 27.08 for the same input, which
reproduces their value to within 0.01 dex.

\subsection*{A.2. Hosts without an X-ray measurement}

For LP~791-18, GJ~3090, TOI-776 and TOI-270 no X-ray luminosity is available and $L_{\rm XUV}$ is
estimated from the M-dwarf X-ray activity--age relationships of \citet{engle24}, which are
constructed for M0--M6.5 dwarfs and calibrated on the age--rotation relationships of
\citet{engle23}. The relations are segmented,
\begin{equation}
\log\!\left(\frac{L_{\rm X}}{L_{\rm bol}}\right) = a\log t + b +
\begin{cases} 0 & \log t < d\\ c\,(\log t - d) & \log t \ge d,\end{cases}
\label{eq:engle}
\end{equation}
with $t$ in Gyr and separate coefficients for the M0--2 and M2.5--6.5 subsets. The saturation phase
ends at $\approx0.47$\,Gyr for the early subset and $\approx2.25$\,Gyr for the mid-late one, and the
post-saturation slopes are $-1.60$ and $-3.40$ respectively. We assign hosts to subsets by spectral
type; K2-18, at M2.5\,V, sits on the boundary, and we have verified that its classification is
unaffected by the choice ($X=0.06$ as mid-late, $0.37$ as early, both far below unity). Because
\citet{engle24} integrate X--UV over 5--1700\,\AA, which is not the band of
Eq.~(\ref{eq:elim}), we take the X-ray relation and pass it through Eq.~(\ref{eq:sf11}) rather than
using the integrated X--UV quantity directly.

We note that the saturated value frequently quoted in this context,
$R_{\rm X,sat}=10^{-3.13}$ \citep{wright11}, is a ratio of X-ray to bolometric luminosity and not of
XUV to bolometric luminosity; given the $L_{\rm EUV}/L_{\rm X}$ ratios above the two differ by more
than an order of magnitude and should not be interchanged.

\subsection*{A.3. The integrated history}

The integrated exposure is derived from Eq.~(\ref{eq:engle}) rather than assumed. Expressed as the
ratio of the time-integrated fluence to the present-day flux times the age, it is 23.3 for the
early-type hosts at 5\,Gyr and 11.5 for the mid-late ones, rising to 62 by 10\,Gyr. Earlier
treatments of this sample parameterised the quantity as a free enhancement factor of 10--50; the
derived values fall inside that range, but they are no longer free, and they differ between hosts
of the same age according to spectral type. We note that the solar-analogue prescription of
\citet{ribas05}, built on a 100\,Myr saturation phase, would substantially understate the
saturation of M dwarfs and is not used here.

\subsection*{A.4. Propagated uncertainty}

The prediction scatter of Eq.~(\ref{eq:sf11}) is approximately 0.25\,dex, a factor of 1.8 in
$L_{\rm XUV}$ and hence, since $\dot M$ is linear in $F_{\rm XUV}$, a factor of 1.8 in $X$. This is
comparable to the range spanned by the heating-efficiency prior, and it enters the Monte Carlo of
Sect.~\ref{sec:4.3} as a fourth log-normal term: 0.25\,dex for hosts with a measured $L_{\rm X}$
and 0.4\,dex for those relying on the activity--age calibration. Where $L_{\rm bol}$ is taken from
published radii and effective temperatures the albedo assumption does not enter; the Monte Carlo
samples a Bond albedo uniform on 0--0.3 only where it does.

Adding these terms widens the intervals in $X$ but does not alter the partition at $X=1$, which is
what the classification uses.

\begin{table*}
\centering
\caption{Inputs to the escape calculation. $\xi=(M_p/3M_\star)^{1/3}(a/R_p)$ is the Roche-lobe
parameter of \citet{erkaev07} and $K$ the corresponding tidal enhancement factor; $\Lambda$ is
computed from Eq.~(\ref{eq:lambda}). Luminosities in erg\,s$^{-1}$, $v_{\rm esc}$ in km\,s$^{-1}$.
Bolometric luminosities of the hosts, used to convert the activity--age relation of
\citet{engle24} into an absolute $L_{\rm X}$, are computed as $4\pi R_\star^2\sigma T_{\rm eff}^4$
from published radii and effective temperatures: LHS~1140, $0.00386\,L_\odot$
($R_\star=0.216\,R_\odot$; \citealt{cadieux24}); LP~791-18, $0.00202\,L_\odot$
($0.171\,R_\odot$; \citealt{crossfield19}); GJ~3090, $0.0384\,L_\odot$
($0.516\,R_\odot$; \citealt{almenara22}); TOI-270, $0.0195\,L_\odot$
($0.378\,R_\odot$; \citealt{vaneylen21}); TOI-776, $0.0310\,L_\odot$
($0.544\,R_\odot$; \citealt{fridlund24}).}
\label{tab:escape}
\begin{tabular}{l r r r r r r r l}
\toprule
Planet & $a$ (au) & $M_\star$ ($M_\odot$) & age (Gyr) & $v_{\rm esc}$ & $\xi$ & $K$ & $\Lambda$ & XUV constraint \\
\hline
LHS 1140 b  & 0.0946 & 0.184 & 5.0  & 20.1 & 40.1 & 0.963 & 108.6 & $L_X=1.34\times10^{26}$ \\
GJ 1214 b   & 0.0150 & 0.178 & 6.0  & 19.6 &  4.7 & 0.684 &  39.1 & $L_X=10^{25.87}$ \\
LP 791-18 c & 0.0296 & 0.139 & 5.0  & 19.1 & 10.6 & 0.859 &  62.1 & activity proxy \\
GJ 3090 b   & 0.0323 & 0.519 & 1.02 & 16.1 &  7.2 & 0.792 &  22.7 & proxy; He outflow \\
TOI-776 c   & 0.1001 & 0.542 & 1--4 & 20.5 & 26.8 & 0.944 &  60.9 & activity proxy; Ly-$\alpha$ escape \\
TOI-270 d   & 0.0721 & 0.386 & 5.0  & 16.8 & 18.4 & 0.919 &  43.9 & activity proxy \\
K2-18 b     & 0.1591 & 0.495 & 2.9  & 20.3 & 37.1 & 0.960 &  98.3 & $L_{\rm XUV}<2.48\times10^{28}$ \\
LTT 3780 c  & 0.0770 & 0.401 & 5.0  & 20.5 & 20.6 & 0.927 &  72.2 & $L_X<5.12\times10^{26}$ \\
\hline
\end{tabular}
\end{table*}

\renewcommand{\thetable}{B\arabic{table}}
\setcounter{table}{0}
\renewcommand{\thefigure}{B\arabic{figure}}
\setcounter{figure}{0}

\section{From metallicity to mean molecular weight}
\label{app:ztomu}

Two planets in the sample are placed above the gas-dwarf threshold by a bound on atmospheric
metallicity rather than by a retrieved $\mu$ (GJ~1214~b and TOI-776~c). Because the mapping from
$Z$ to $\mu$ is not unique we set it out explicitly and quote $\mu$ as a range over the speciation
cases. GJ~3090~b is not included here: its metallicity and $\mu$ are joint outputs of a single
retrieval containing a free cloud deck \citep{parker25}, so the conversion below does not apply to
it.

We compute $\mu$ elementally. Protosolar abundances by number relative to hydrogen are
He/H\,$=8.51\times10^{-2}$, O/H\,$=4.90\times10^{-4}$, C/H\,$=2.69\times10^{-4}$ and
N/H\,$=6.76\times10^{-5}$. Heavy elements are scaled by the metallicity factor $Z$ in solar units
with the He/H ratio held at its protosolar value. Carbon is assigned to CH$_4$, CO or CO$_2$;
nitrogen to N$_2$ or NH$_3$; oxygen not consumed by carbon monoxide or dioxide forms H$_2$O.
Hydrogen not consumed in hydride formation remains as H$_2$. The mean molecular weight is the total
mass divided by the total particle number. At $Z=1$ the calculation returns $\mu=2.33$\,amu, as
required.

Table~\ref{tab:ztomu} gives the result. The spread between reducing and oxidised speciation is
modest at low metallicity but reaches a factor of $\approx1.4$ by $Z\sim300$, and it is this
spread, not the measurement uncertainty on $Z$, that dominates the error on $\mu$ for the bounded
planets.

Two features bear on the argument of Sect.~\ref{sec:5.2}. The reducing cases become unrealisable
above $Z\approx400$: the hydrogen budget is exhausted forming CH$_4$ and H$_2$O, so a
methane-bearing atmosphere cannot exceed $\mu\approx11$\,amu at any metallicity, and the upper part
of the hybrid range is accessible only to oxidised envelopes. And the metallicity required to cross
the gas-dwarf boundary is $Z\approx66$--$80$, well below the values inferred for either bounded
planet, so their placement above the threshold does not depend on the speciation adopted.

As an external check, LP~791-18~c has both a retrieved $\mu$ of $7.68^{+2.24}_{-1.48}$\,amu and an
inferred $Z\approx316\,Z_\odot$ \citep{roy25}. The calculation returns $\mu=7.6$--$10.5$\,amu at
that metallicity, consistent with the retrieval.

\begin{table}
\centering
\caption{Mean molecular weight as a function of atmospheric metallicity and assumed speciation.
Entries marked \ldots are not realisable: the hydrogen budget is exhausted by hydride formation at
that metallicity.}
\label{tab:ztomu}
\begin{tabular}{r c c c c}
\toprule
$Z/Z_\odot$ & CH$_4$/N$_2$ & CH$_4$/NH$_3$ & CO$_2$/N$_2$ & CO/N$_2$ \\
\hline
   1 &  2.33 &  2.33 &  2.32 &  2.32 \\
  10 &  2.52 &  2.52 &  2.50 &  2.50 \\
  30 &  2.96 &  2.97 &  2.88 &  2.88 \\
 100 &  4.54 &  4.60 &  4.14 &  4.14 \\
 150 &  5.73 &  5.84 &  5.00 &  5.00 \\
 180 &  6.47 &  6.62 &  5.49 &  5.49 \\
 240 &  8.01 &  8.26 &  6.44 &  6.44 \\
 316 & 10.08 & 10.51 &  7.56 &  7.56 \\
 500 & \ldots & \ldots &  9.99 &  9.99 \\
1000 & \ldots & \ldots & 15.05 & 15.05 \\
3000 & \ldots & \ldots & 25.04 & \ldots \\
\hline
\end{tabular}
\end{table}

\end{document}